\documentstyle[prb,aps]{revtex}

\begin{document}
\draft
\title{Magnetic Pinning of Vortices in a Superconducting Film: The
(anti)vortex-magnetic dipole interaction energy in the London approximation}
\author{M. V. Milo\v{s}evi\'{c}, S. V. Yampolskii,\cite{perm} and F. M. Peeters\cite
{aut2}}
\address{Departement \ Natuurkunde, Universiteit Antwerpen (UIA), \\
Universiteitsplein 1, \ B-2610 Antwerpen, Belgium}
\date{\today}
\maketitle

\begin{abstract}
The interaction between a superconducting vortex or antivortex in a
superconducting film and a magnetic dipole with in- or out-of-plane
magnetization is investigated within the London approximation. The
dependence of the interaction energy on the dipole-vortex distance and the
film thickness is studied and analytical results are obtained in limiting
cases. We show how the short range interaction with the magnetic dipole
makes the co-existence of vortices and antivortices possible. Different
configurations with vortices and antivortices are investigated.
\end{abstract}


\section{Introduction}

The value of the critical current is one of the decisive factors for the
usefulness of a superconductor (SC). Large values of the critical currents
in superconductors are usually obtained through pinning of vortices to
different inhomogeneities in a superconductor. In this respect, external and
internal surfaces can be treated as inhomogeneities.~\cite{shmidt} Pinning
centra are introduced e.g. by locally destroying the crystal structure
through e.g. bombardment with high energy particles. Recently,
nanostructuring of a superconducting film with a regular area of holes~\cite
{baert} has lead to a large increase of the critical current, in particular
at the so-called matching magnetic fields. An alternative approach is to
deposit an area of ferromagnetic dots near a superconducting film which acts
as very effective trapping centra for the vortices. Recently, it was
predicted,~\cite{bul} although not verified, that an increase of the pinning
effects by two orders of magnitude can be realized. After substantial
progress in the preparation of regular magnetic arrays on superconductors~ 
\cite{schuller} and considering the importance of such structures for
magnetic device and storage technologies, these hybrid systems became very
interesting both from a theoretical and an experimental point of view.
Macroscopic pinning phenomena have already been explored experimentally,~ 
\cite{schuller,leuven1,leuven2} but a theoretical analysis of the magnetic
and superconducting response in such systems is still lacking. In the
majority of recently proposed models~\cite{lip1,lip2,kay1,kay2,hwa} the
inhomogeneous magnetic field of a ferromagnet induces screening currents in
a SC, which, in turn, generate a magnetic field influencing the applied
field. Consequently, this problem must be solved self-consistently.
Furthermore, the finite thickness of both the superconductor and the
ferromagnetic material was not taken into account in previous theoretical
treatments.

Other theoretical studies involving finite size ferromagnets were mainly
restricted to the problem of a magnetic dot with out-of-plane magnetization
embedded in a superconducting film.~\cite{marm,fertig} Marmorkos {\it et al.}%
~\cite{marm} solved the non-linear Ginzburg-Landau (GL) equation
numerically, with appropriate boundary conditions for an infinitely long
ferromagnetic cylinder penetrating the superconducting\ film, and found a
correspondence between the value of the magnetization and the vorticity of
the most energetically favorable giant-vortex state. At that time, the
interaction between a small ferromagnetic particle, which may be considered
as a point magnetic dipole (MD), and a type II superconductor was of
interest,~\cite{xu,wei} and the magnetostatic energy and levitation force
acting on a dipole were calculated, but still within a thin film
approximation. In experiments,~\cite{mark} the magnetic field of the MD was
supposed to be weak, and not able to drastically change the structure of the
superconducting state in the sample. Obviously, the MD could create
additional vortices near the surface, and this process can be described
theoretically as proposed in Ref.~\cite{wei}, by a simple comparison of the
free energies of the system with and without vortex. However, the
spontaneous creation of a vortex-antivortex pair as a possible lower energy
state was never considered.

Motivated by recent experiments,~\cite{morgan,bael} we focus in this paper
on a model system consisting of a type II thin superconducting film (SC) and
a magnetic dipole placed above (below) it which acts as a pinning center
(Fig.~\ref{fig1}). We study in detail, using the London approximation, how
the system is perturbed in the neighborhood of the dipole. The
superconducting film lies in the $z=0$ xy-plane while the MD is positioned
at $(x,y,z)=(0,0,a)$, and is magnetized in the positive z- or x-axis
direction. To avoid the proximity effect and exchange of electrons between
MD and SC we suppose a thin layer of insulating oxide between them as is
usually the case in the experiment.

The paper is organized as follows. In the next section we present the
general formalism. In Sec.~III, we discuss the pinning potential of the
magnetic dipole. In Sec.~IV the total interaction energy in the system is
calculated in the presence of vortex-antivortex pairs, and their most
favorable position is determined. The question of stability of such vortex
configurations is then analyzed and our conclusions are given in Sec.~V.

\section{Theoretical formalism}

We consider a magnetic point dipole with magnetization $\vec{M}$, placed
outside a type II SC film interacting with a single vortex in the SC. Within
the London approximation, the total energy of the stationary
magnet-superconductor system is given by~\cite{wei}

\begin{equation}
\vec{F}=\frac{1}{8\pi }\int dv_{1}\left[ \vec{h}^{2}+\lambda ^{2}(rot\vec{h}%
)^{2}\right] +\frac{1}{8\pi }\int dv_{2}\vec{h}^{2}\text{,}  \eqnum{1}
\label{free1}
\end{equation}
where $\lambda $ is the penetration depth and $\vec{h}$ is the total field
present in the system. The integral of the first term is taken over the
volume of the SC, and the integral of the second term is taken outside,
except for the volume of the dipole. Therefore, we may rewrite Eq.~(\ref
{free1}) as 
\begin{eqnarray*}
F &=&\frac{1}{8\pi }\int dV^{(i)}\left[ \vec{h}^{2}+\lambda ^{2}(rot\vec{h}%
)^{2}\right] +\frac{1}{8\pi }\int dV^{(o)}\vec{h}^{2}-\frac{1}{8\pi }\int
dV^{(md)}\vec{h}^{2} \\
&=&\frac{1}{8\pi }\int dV\left[ rot\vec{h}\cdot \left( \vec{A}+\lambda
^{2}rot\vec{h}\right) \right] -\frac{1}{8\pi }\int d\vec{S}\cdot \left( \vec{%
h}\times \vec{A}\right) -\frac{1}{8\pi }\int dV^{(md)}\vec{h}^{2},
\end{eqnarray*}
where we integrate over the volume inside $V^{(i)}$ and outside $V^{(o)}$
the superconductor, while $V^{(md)}$ denotes the volume of the magnetic
dipole. We choose the surface $S$ far away from the superconductor where we
can apply the boundary condition $\vec{h}\rightarrow 0$. Due to the London
equation, the field of the magnetic dipole satisfies 
\begin{equation}
\vec{h}_{m}+\lambda ^{2}rot\left( rot\vec{h}_{m}\right) =0\text{.}  \eqnum{2}
\end{equation}
The magnetic field and corresponding vector potential can be written as $%
\vec{h}=\vec{h}_{m}+\vec{h}_{v}$, $\vec{A}=\vec{A}_{m}+\vec{A}_{v}$ where
indexes $v$ and $m$ refer to the vortex and ferromagnet, respectively. Now,
from Eq.~(\ref{free1}) we obtain the vortex-dipole interaction energy as 
\[
F_{mv}=\frac{1}{8\pi }\int dV\left[ rot\vec{h}_{m}\cdot \left( \vec{A}%
_{v}+\lambda ^{2}rot\vec{h}_{v}\right) \right] =\frac{1}{2c}\int dV^{(i)}%
\left[ \vec{j}_{m}\cdot \vec{\Phi}_{v}\right] +\frac{1}{2}\int dV_{v}^{(o)}%
\vec{A}\cdot rot\vec{M}-\int dV_{v}^{(md)}\vec{h}\cdot \vec{M} 
\]
which after simple integral transformations becomes 
\begin{equation}
F_{mv}=\frac{1}{2c}\int dV^{(i)}\left[ \vec{j}_{m}\cdot \vec{\Phi}_{v}\right]
-\frac{1}{2}\int dV^{(md)}\vec{h}_{v}\cdot \vec{M}\text{,}  \eqnum{3}
\label{freemv}
\end{equation}
where $\vec{\Phi}_{v}=\left( \Phi _{\rho },\Phi _{\varphi },\Phi _{z}\right)
=\left( 0,\Phi _{0}/(2\pi \rho ),0\right) $ denotes the vortex magnetic flux
vector ($\Phi _{0}$ is the flux quantum), and $\vec{M}$ is the magnetization
of the dipole. As one can see, the interaction energy in this system
consists of two parts, namely, (i) the interaction between the Meissner
currents generated in the SC ($\vec{j}_{m}$) by the MD and the vortex, and
(ii) the interaction between the vortex magnetic field and the MD.

In order to obtain the value of the current induced in the superconductor\
by a magnetic dipole with moment $\vec{m}$ ($\vec{M}=\vec{m}\delta (x)\delta
(y)\delta (z-a)$), we solve the corresponding equation for the vector
potential~\cite{mel} 
\begin{equation}
rot\left( rot\vec{A}_{m}\right) +\frac{1}{\lambda ^{2}}\Theta \left(
d/2-\left| z\right| \right) \vec{A}_{m}=4\pi rot\left( \vec{m}\delta
(x)\delta (y)\delta (z-a)\right) ,  \eqnum{4}  \label{cur}
\end{equation}
where $\delta (...)$ is the Dirac delta function. The results for both
vector potential and magnetic field, for different orientations of the
magnetic moment of the dipole, are given in the Appendix.

\section{Magnetic dipole - vortex interaction energy}

The superconducting current induced in an infinite superconducting film with
thickness $d$ ($-\frac{d}{2}<z<\frac{d}{2}$) by a magnetic dipole with {\it %
out-of-plane magnetization}, i.e. $\vec{m}=m\vec{e}_{z}$ located at $z=a$ is
obtained as a solution of Eq.~(\ref{cur}) which is given by 
\begin{equation}
j_{\varphi }(\rho ,z)=-\frac{cm\Phi _{0}}{2\pi \lambda ^{3}}%
\int\limits_{0}^{\infty }dq\exp \left\{ -q\left( \left| a\right| -\frac{d}{2}%
\right) \right\} q^{2}J_{1}(q\rho )C\left( q,z\right) \text{,}  \eqnum{5a}
\label{curop1}
\end{equation}
with 
\begin{equation}
C\left( q,z\right) =\frac{k~cosh(k(\frac{d}{2}+z))+q~sinh(k(\frac{d}{2}+z))}{%
(k^{2}+q^{2})~sinh(kd)+2kq~cosh(kd)}\text{,}  \eqnum{5b}  \label{curop}
\end{equation}
where $k=\sqrt{1+q^{2}}$, $\rho =\sqrt{x^{2}+y^{2}}$, $sinh$ and $cosh$
denote the hyperbolic trigonometric functions and $J_{v}(x)$ is the Bessel
function. For a MD placed under the SC, one should replace $z$ by $-z$ in
Eq.~(\ref{curop}). The magnetic moment of the dipole is measured in units of 
$m_{0}=\Phi _{0}\lambda $, and all distances are scaled in units of $\lambda 
$. These units will be used in the rest of the paper. Clearly, Eq.~(\ref{cur}%
) can also be used for a film of finite dimensions $L$ in the limits $\rho
,a<<L$. The above integral~(\ref{curop1}) can be solved analytically for
certain asymptotic limits 
\begin{equation}
j_{\varphi }(\rho ,z)=\left\{ 
\begin{array}{l}
-\frac{3cm\Phi _{0}\rho l}{2\pi \lambda ^{3}\left( \rho ^{2}+l^{2}\right)
^{5/2}}\frac{\cosh (d/2+z)}{\sinh (d)}\ \ \ \ \ \ \ \ \ \ \ \ \ ,~\rho
>>\max (l,1/d)~\text{or}~l>>\max (\rho ,1/d), \\ 
-\frac{cm\Phi _{0}}{4\pi \lambda ^{3}}\left[ \frac{\rho }{\left( \rho
^{2}+l^{2}\right) ^{3/2}}+\frac{d\left( \sqrt{\rho ^{2}+l^{2}}-l\right) }{%
2\rho \sqrt{\rho ^{2}+l^{2}}}\right] ,~\rho ,l<\frac{1}{d},~d<<1,
\end{array}
\right.  \eqnum{6}  \label{asympj}
\end{equation}
where $l=\left| a\right| -\frac{d}{2}$ is the distance between the MD and
the top surface of the SC film.

Inserting the well known expression for the magnetic field of a vortex (see
Eq.~(A4) in the Appendix), into Eq.~(\ref{freemv}) we find 
\begin{equation}
F_{mv}=\frac{m\Phi _{0}^{2}}{2\pi \lambda }U_{\perp }\left( \rho _{v}\right) 
\text{,}  \eqnum{7a}  \label{fmvop}
\end{equation}
where $\rho =\rho _{v}$ denotes the position of the vortex, and 
\begin{equation}
U_{\perp }\left( \rho _{v}\right) =-\int\limits_{0}^{\infty }dq\frac{q}{%
k(k+q\coth (kd/2))}J_{0}\left( q\rho _{v}\right) \exp (-ql)\text{.} 
\eqnum{7b}
\end{equation}
In some limiting cases it is possible to solve this integral analytically:

(1) for $d<<1$ and $\rho _{v}<1/d$ we found: $U_{\perp }\left( \rho
_{v}\right) \approx -d\big/(2\left( \rho _{v}^{2}+l^{2}\right) ^{1/2})$;

(2) for $d<1$ and a) $\rho _{v}>a$: $U_{\perp }\left( \rho _{v}\right)
\approx -d\big/(2\left( \rho _{v}^{2}+l^{2}\right) ^{1/2})+\pi d^{2}\left[
H_{0}(\rho _{v}d/2)-Y_{0}(\rho _{v}d/2)\right] \big/8$,

\qquad \qquad \qquad \qquad b) $\rho _{v}>>1/d$: $U_{\perp }\left( \rho
_{v}\right) \approx -2\big/(d\rho _{v}^{3})$;

(3) for $d>1$ and $\rho _{v}>>\max (l,1/d)~$or$~l>>\max (\rho _{v},1/d)$ we
obtained: $U_{\perp }\left( \rho _{v}\right) \approx -l\big/\left( \rho
_{v}^{2}+l^{2}\right) ^{3/2}$.

\noindent Here, $H_{v}(x)$ and $Y_{v}(x)$ denote the Struve and Bessel
function, respectively. One should notice that our asymptotic results for a
thin SC film differ from those given in Eq.~(7) of Ref.~\cite{hwa} by a
factor of two, and are in agreement with the calculations of Ref.~\cite{wei}
(Eq.~(3.16) of Ref.~\cite{wei}).

In Fig.~\ref{fig2}(a,b) the interaction energy as function of the position
of the vortex are shown, for different \ (a) vertical positions of the
magnetic dipole, and (b) thickness of the SC. Please note that the
asymptotic expressions for an extremely thin SC, namely $d<<1$, give a very
good description of the interaction potential (see inset of Fig.~\ref{fig2}%
(b)). As expected, the problem of the interaction energy in this system is
axially symmetric and this is illustrated in Fig.~\ref{fig2}(c). In Fig.~\ref
{fig2}(a), we notice that the energy curves for different vertical positions
of the dipole cross each other for large MD-vortex distances, for the case
of a thick SC, namely $d>1$. Consequently, for a MD which is higher above
the SC film, interaction with the vortex is stronger at large distances,
i.e. $\rho _{v}>>1/d$ (see our approximate results above, case (3)). This
can be understood as follows. In the case of a thick superconductor, the
magnetic field does not penetrate through the SC, and the component of the
field tangential to the surface becomes important. When the MD approaches
the SC surface, the value of this field component at large distances from
the dipole increases, as shown in the inset of Fig.~\ref{fig2}(a).
Therefore, the interaction between the dipole and the vortex is stronger for
smaller $l$ when the MD-vortex distance $\rho _{v}$ is small, and opposite,
for large $\rho _{v}$, the interaction energy grows if $l$ is larger.

Naturally, the minimum of the interaction energy depicts the energetically
favorable position of the vortex. Therefore, from Fig.~\ref{fig2} one
notices that for parallel alignment of the MD magnetization and the vortex
the interaction will be purely attractive, and the vortex is pinned under
the MD, independently of the parameters of the system (thickness of the SC,
vertical position of the MD and its magnetization strength, etc.). These
parameters only determine the strength of this attractive interaction. For
the anti-parallel orientation of the magnetic moment, the vortex is repelled.

Following the same procedure, for a MD with {\it in-plane magnetization} ($%
\vec{m}=m\vec{e}_{\rho }$, for $\varphi =0$, and $\vec{m}=0$ for $\varphi
\neq 0$) we have 
\begin{equation}
F_{mv}=\frac{m\Phi _{0}^{2}}{2\pi \lambda }U_{\parallel }\left( \rho
_{v},\varphi \right) \text{,}  \eqnum{8a}  \label{fmvip}
\end{equation}
where $\varphi $ is the angle in the SC plane between the direction of the
magnetic moment of the dipole and the point of interest, and 
\begin{equation}
U_{\parallel }\left( \rho _{v},\varphi \right) =\int\limits_{0}^{\infty }dq%
\frac{q\cos \varphi }{k(k+q\coth (kd/2))}J_{1}\left( q\rho _{v}\right) \exp
(-ql),  \eqnum{8b}
\end{equation}
for which we obtain the following analytic asymptotic results 
\begin{equation}
U_{\parallel }\left( \rho _{v},\varphi \right) =\left\{ 
\begin{array}{l}
\frac{\rho _{v}\cos \varphi }{\left( \rho _{v}^{2}+l^{2}\right) ^{3/2}}%
\qquad \qquad \ ,~\rho >>\max (l,1/d)~\text{or}~l>>\max (\rho ,1/d)\text{,}%
~d>1, \\ 
\frac{\rho _{v}d\cos \varphi }{2\sqrt{\rho _{v}^{2}+l^{2}}\left( l+\sqrt{%
\rho _{v}^{2}+l^{2}}\right) },~d<<1.
\end{array}
\right.  \eqnum{9}
\end{equation}
In Fig.~\ref{fig3}(a,b), the energy is plotted as function of the position
of the vortex when it is moved along the direction of the MD magnetization ($%
\varphi =\pi $), and Fig.~\ref{fig3}(c) shows the complete contour plot. The
vortex is attracted to one side (where the magnetic field of the MD
penetrates the SC in the same direction as the vortex, see Fig.~\ref{fig1}%
(b)) and repelled on the opposite side of the MD. Moreover, it is pinned in
a spot whose position does not depend on the magnetization of the MD but it
is influenced by the vertical position of the dipole and the thickness of
the SC. When the dipole approaches the SC or the thickness of the
superconductor is increased, the interaction becomes stronger and the vortex
will move closer to the MD.

From our asymptotic expressions for the interaction energy the most
favorable position of the vortex can be easily obtained by minimizing the
energy. In the thin film approximation ($d<<1$), we obtain $\rho _{v}^{\ast
}=\sqrt{2+2\sqrt{5}}l\big/2\approx 1.272~l,$ while for $d>1$, but $l>>1/d$
we have $\rho _{v}^{\ast }=l\big/\sqrt{2}\approx 0.7071~l$. In Fig.~\ref
{fig4}(a) we show the $d-$dependence of the equilibrium position as function
of the thickness of the SC. The $\rho _{v}^{\ast }(l)$ function is in
general non-linear, as shown in Fig.~\ref{fig4}(b). A peculiar fact is that
for $d>2$, the energetically preferable position of the vortex depends only
on $l$. This follows from the fact that the magnetic field of the dipole
penetrates only in the surface layer of the superconductor down to the
penetration depth $\lambda $ so that further increase of the SC thickness
does not affect the total interaction. From Fig.~\ref{fig4}(b) one should
notice that our asymptotic results describe nicely the behavior of the
system for $d<<1$ or $l>>d$, when $d>1$.

For the case of in-plane MD, it should be emphasized that the vortex is
attracted to the side of the MD where the magnetic field is parallel to the
field of the vortex, which is similar to the case of the out-of-plane
magnetized dipole. This conclusion also follows from the observation of the
interaction between the induced currents and the vortex. For in-plane
magnetized MD above the superconductor, the components of the
superconducting current in the absence of vortices are given by 
\begin{equation}
j_{\rho }(\rho ,\varphi ,z)=-\frac{cm\Phi _{0}\sin \varphi }{2\pi \lambda
^{3}\rho }\int\limits_{0}^{\infty }dq\exp (-ql)qJ_{1}(q\rho )C(q,z)\text{,} 
\eqnum{10a}
\end{equation}
\begin{equation}
j_{\varphi }(\rho ,\varphi ,z)=\frac{cm\Phi _{0}\cos \varphi }{2\pi \lambda
^{3}}\int\limits_{0}^{\infty }dq\exp (-ql)q^{2}\left( \frac{J_{1}(q\rho )}{%
q\rho }-J_{0}(q\rho )\right) C(q,z)\text{,}  \eqnum{10b}
\end{equation}
\begin{equation}
j_{z}(\rho ,\varphi ,z)=0.  \eqnum{10c}
\end{equation}
We obtained the following asymptotic behavior of these components:

a) $\rho >>\max (l,1/d)~$or$~l>>\max (\rho ,1/d)$%
\begin{eqnarray}
j_{\rho }(\rho ,\varphi ,z) &=&-\frac{cm\Phi _{0}\sin \varphi }{2\pi \lambda
^{3}\left( \rho ^{2}+l^{2}\right) ^{3/2}}\frac{\cosh (d/2+z)}{\sinh (d)}, 
\eqnum{11a} \\
j_{\varphi }(\rho ,\varphi ,z) &=&\frac{cm\Phi _{0}\cos \varphi }{2\pi
\lambda ^{3}\left( \rho ^{2}+l^{2}\right) ^{3/2}}\frac{2\rho ^{2}-l^{2}}{%
\rho ^{2}+l^{2}}\frac{\cosh (d/2+z)}{\sinh (d)},  \eqnum{11b}
\end{eqnarray}

b) $\rho ,l<\frac{1}{d},~d<<1$%
\begin{eqnarray}
j_{\rho }(\rho ,\varphi ,z) &=&-\frac{cm\Phi _{0}\sin \varphi }{4\pi \lambda
^{3}}\frac{\sqrt{\rho ^{2}+l^{2}}-l}{\rho ^{2}\sqrt{\rho ^{2}+l^{2}}}, 
\eqnum{11c} \\
j_{\varphi }(\rho ,\varphi ,z) &=&\frac{cm\Phi _{0}\cos \varphi }{4\pi
\lambda ^{3}}\left[ \frac{\sqrt{\rho ^{2}+l^{2}}-l}{\rho ^{2}\sqrt{\rho
^{2}+l^{2}}}-\frac{l}{\left( \rho ^{2}+l^{2}\right) ^{3/2}}\right] . 
\eqnum{11d}
\end{eqnarray}
For the case of the MD under the SC, these currents change sign. The
detailed expressions for the magnetic fields and the vector potential as
solution of Eq.~(\ref{cur}) are given in the Appendix. The vector plots of
the screening currents for both directions of the MD magnetization are
presented in Fig.~\ref{fig5}.

Previous theoretical studies~\cite{aut} on the interaction of a magnetic
moment and a vortex used a Gibbs free energy argument to show that this
force was attractive. In our case, the pinning force between a vortex and
the MD consists of two parts: (i) the interaction between the vortex
fringing field and the magnetic moment, and (ii) the interaction between the
vortex and the screening currents created in response to the magnetic dipole
field. One should notice (Fig.~\ref{fig5}(a)) that although the magnetic
field of a MD with out-of plane magnetization changes sign in the SC plane,
the superconducting screening current is always in the same direction, and
therefore, the interaction between the MD and a vortex with parallel
orientation of the field is attractive (consequently, the vortex will sit
under the MD), and analogously, for anti-parallel orientation of the
magnetic fields they repel each other.

For the case of a MD with in-plane magnetization, the Meissner current
vector plot in Fig.~\ref{fig5}(b) shows that the direction of the current is
such that the vortex is attracted to the region where the magnetic field of
a MD is parallel to it and repelled on the other side of a MD due to the
opposite direction of the current.

\section{Pinning of multiple vortices and/or anti-vortices}

\subsection{In-plane magnetized dipole}

Obviously, in the case of in-plane magnetization, the vortex will be
repelled on the side of the MD to which the magnetic moment points to and
pinned on the opposite side (and vice versa for the antivortex). The case
when the dipole itself creates vortices was experimentally investigated by
Van Bael {\it et al.}~\cite{bael}, where it was found that the stray field
of the in-plane magnetic dipole induces a vortex-antivortex pair (VA) at the
poles, at positions predicted by our analysis (vortex towards negative pole,
antivortex towards positive pole). Therefore, this asymmetric pinning
potential provides stability for {\it vortex-antivortex} configurations. One
would expect that the position of the pinning sites depends on the magnetic
moment of the MD and the number of flux quanta carried by each vortex. To
determine the preferable position of the vortex-antivortex pair in the
presence of an in-plane MD we put the vortex at ($\rho _{v},\varphi _{v}$)
and the antivortex at ($\rho _{av},\varphi _{av}$) with respect to the MD
and calculate their interaction with the dipole. The vortex-antivortex
interaction energy is given by~\cite{ita} 
\begin{equation}
F_{v-av}(R,L_{v},L_{av})=-\frac{L_{v}L_{av}\Phi _{0}^{2}}{8\pi ^{2}\lambda }%
\left( dK_{0}\left( R\right) +2\int\limits_{0}^{\infty }dq\frac{J_{0}(qR)}{%
k^{2}Q}\right) \text{,}  \eqnum{12}  \label{vav}
\end{equation}
where $R=\sqrt{\rho _{v}^{2}+\rho _{av}^{2}-2\rho _{v}\rho _{av}cos(\varphi
_{v}-\varphi _{av})}$ is the distance between the vortex and antivortex, $%
K_{0}\left( x\right) $ is the MacDonald function, and $L_{v}$, $L_{av}$ are
the vorticity of the vortex and anti-vortex, respectively.

First, let us suppose that the vortex-antivortex pair appears due to the
stray field of the dipole. In that case, this pair nucleates where the
superconducting current is maximal, namely, under the dipole (Fig.~\ref{fig5}%
(b)), with vortex and antivortex situated on opposite sides of the dipole.
Therefore, due to the symmetry of the applied potential (Fig.~\ref{fig3}%
(c)), we may assume that $\rho _{v}=\rho _{av}$, and $\varphi _{v}=\pi
-\varphi _{av}$. To investigate the stability of such a pair of vortices, we
calculate the total interaction energy in this system with respect to the
position of the vortices. The results are shown as contourplots in Figs.~\ref
{fig6}(a-c) for different values of the magnetic moment of the dipole. The
lowest value of the interaction energy gives the energetically favorable
position of the vortex-antivortex pair [apart from the global minimum at $%
x=0 $ (zero distance between vortices), where the energy equals $-\infty $].
Obviously, the lateral motion of vortices is forbidden by the high energy
barriers, while along the direction of the dipole, also an energy barrier
exists, but lower than the lateral ones, showing us the preferable direction
of possible vortex-antivortex annihilation.

Knowing this, we put the vortex and antivortex along the direction of the
MD, in such a way that $\rho _{v}=\rho _{av}$, and $\varphi _{v}=\pi $, $%
\varphi _{av}=0$, and investigate the pinning potential as function of the
distance of the vortex and antivortex from the MD. We substitute $R=2\rho
_{v}$ in Eq.~(\ref{vav}) and add this energy to the values of the MD-vortex
and MD-antivortex attraction energy obtained from Eq.~(\ref{fmvip}). The
result is shown in Fig.~\ref{fig7} for (a) different values of the vorticity
($L_{v}$, $L_{av}$), and (b) different magnetization and $L_{v}=L_{av}=1$.
After minimization of the total energy over $\rho _{v}$, we obtain the
position of the vortex-antivortex pair. In the thin superconductor case ($%
d<1 $) this leads to the non-linear equation

\begin{equation}
\frac{2\pi m\rho _{v}}{\sqrt{l^{2}+\rho _{v}^{2}}\left( l+\sqrt{l^{2}+\rho
_{v}^{2}}\right) }\left[ 1-\rho _{v}^{2}\left( \frac{1}{l^{2}+\rho _{v}^{2}}+%
\frac{1}{\sqrt{l^{2}+\rho _{v}^{2}}\left( l+\sqrt{l^{2}+\rho _{v}^{2}}%
\right) }\right) \right] =\frac{L_{v}L_{av}}{L_{v}+L_{av}}\text{.} 
\eqnum{13}  \label{vort}
\end{equation}

Although all forces acting in the system are attractive, it is clearly
visible that annihilation of vortices can be prevented by their strong
confinement at the poles of the MD (the position of vortex-antivortex pairs
is illustrated by the open circles in Fig.~\ref{fig7}). The interaction
energy shows an absolute minimum for $\rho _{v}=0$, implying the
annihilation, but in order to do that vortices have to cross a potential
barrier. Therefore, the vortex-antivortex configuration can be stable in
this system. To determine if this vortex state is metastable or the ground
state one should compare its total energy\ to the one without vortices. In
that case, the existence of the pairs as the ground state will not depend
only on the magnetization of the dipole but also on the parameters of the
superconductor (i.e. $\lambda ,\xi $) which influence the equilibrium phase
diagram via the self-energy of the vortices.

From Fig.~\ref{fig7}(b) it is clear that in the case of $L_{v}=L_{av}=1$
there exists a critical MD magnetization for which the annihilation barrier
disappears. More generally, the critical value of the magnetic moment of the
MD, in the thin SC film limit, follows from Eq.~(\ref{vort}) (by the
condition that the function has no solution) and is given by 
\begin{equation}
m^{\ast }=1.08529~\frac{L_{v}L_{av}l}{L_{v}+L_{av}}.  \eqnum{14}
\label{prag1}
\end{equation}
From this equation the critical values of the vorticity, when attraction
becomes stronger than the pinning, can also be estimated. For $L_{v}=L_{av}$
we have $m^{\ast }=1.08529L_{v}l/2$ and using the fact that $\Phi ^{+}/\Phi
_{0}=2m/l$ for an in-plane MD we obtain the critical condition $\Phi
^{+}/\Phi _{0}=1.08529L_{v}$, where $\Phi ^{+}$ denotes the flux through the
region of the positive stray field of the dipole at the SC surface. This
implies that there is a connection between the appearance of stable
vortex-antivortex pairs and the amount of penetrating flux of the magnetic
dipole. Notice that the critical $\Phi ^{+}$ is not exactly quantized in
units of $\Phi _{0}$ which is a mesoscopic effect.~\cite{ben} The
quantization condition $\Phi =\oint_{C}\vec{A}\cdot d\vec{l}=L\Phi _{0}$
cannot be used because it is not possible to construct a contour $C$ around
the positive stray field region where the current is zero (see Fig.~\ref
{fig5}(b)). The inhomogeneous magnetic field of the dipole stimulates the
creation of a vortex and an antivortex at opposite poles. These vortices
cannot be treated independently and it turns out that larger flux is needed
in order not only to create this pair but also to keep them apart.

As shown above, for the case of the experimentally important thin SC film
approximation, the position of the vortex-antivortex pair with respect to
the dipole can be obtained from Eq.~(\ref{vort}). This position depends both
on the vertical position of the dipole and its magnetization strength. For
fixed magnetic moment of the dipole, the vortex-antivortex pair can only be
stabilized for $l$ below some critical value which is obtained from Eq.~(\ref
{prag1}). If we put $m=m^{\ast }$ in Eq.~(\ref{vort}), we obtain the simple
dependence $\rho _{v}^{\ast }=0.4904l^{\ast }$. The dependence of $\rho
_{v}^{\ast }$ on $l$ is shown in Fig.~\ref{fig8}, for different values of
the magnetization, where also the stability region of the vortex-antivortex
pair is indicated. One can see three regions in this diagram: i) the region
where the vortex-antivortex pair is unstable, which is bounded by $\rho
_{v}^{\ast }=0.4904l$, for any value of the magnetization of the MD, ii) the
region of stability, and iii) the forbidden area, for $\rho _{v}^{\ast
}>1.272l$. The latter condition follows from Sec. III (see Fig.~\ref{fig4}%
(b)). Notice that with increasing distance between the dipole and the SC
film, the pinning sites move further from the center of the dipole, up to a
certain point when the interaction between vortices overwhelms the pinning
force. After that, the distance between vortex and antivortex decreases and
they finally annihilate. The maximum distance $\rho _{v}^{\ast }$\ follows
from Eq.~(\ref{vort}) by taking the derivative with respect to $l$ and leads
to $\rho _{v~max}^{\ast }=0.7071l_{max}$ (open dots in Fig.~\ref{fig8}),
which again corresponds to the single vortex situation for $l>>d$ (see Fig.~%
\ref{fig4}).

\subsection{Out-of-plane magnetized dipole}

Due to the strong field inhomogeneity and the reversal of the direction of
the magnetic field in the vicinity of the MD with out-of plane
magnetization, and the fact that the net magnetic flux due to the MD in the
SC plane equals zero, one expects that vortex-antivortex configurations
might be stable in such applied field, as predicted earlier in Ref.~\cite
{misko} for a finite size superconductor. As shown in Fig.~\ref{fig2}, an
individual vortex is strongly attracted by the MD and the antivortices are
repelled. Therefore we artificially put one vortex with vorticity $L_{v}$
under the MD and assume the existence of a ring of $n_{av}$ single
antivortices around it, with radius $\rho _{v}$.~\cite{misko} Adding the
interaction between each two vortices to our previous expression for the
interaction energy, we obtain the total interaction energy 
\begin{equation}
F_{int}=\frac{m\Phi _{0}^{2}}{2\pi \lambda }\left( L_{v}U_{\perp }\left(
0\right) -n_{av}U_{\perp }\left( \rho _{v}\right) \right)
+n_{av}F_{v-av}(\rho _{v},L_{v},1)-\frac{n_{av}}{2}%
\sum_{j=1}^{n_{av}-1}F_{v-av}(\rho _{v}\sqrt{2-2cos\frac{2\pi j}{n_{av}}}%
,1,1),  \eqnum{15}  \label{inten}
\end{equation}
where the first two terms describe the dipole-vortex and dipole-antivortex
interactions, the third term is the vortex-antivortex attraction term, and
the last term is the repulsion energy between antivortices. The function $%
F_{v-av}$ is given by Eq.~(\ref{vav}). Although the vortex is attracted by
the antivortices, annihilation is prevented by the repulsion between the
antivortex and the MD (for example, see Fig.~\ref{fig9}, for $L_{v}=n_{v}=1$%
). Naturally, this energy barrier becomes smaller with decreasing magnetic
moment of the dipole. Nevertheless, if a barrier exists, the interaction
energy shows one local minimum, meaning that antivortices would not be
repelled to infinity but to a certain point. The position of this local
minimum we obtained in the same manner as previously, which within the thin
film approximation, leads to the equation 
\begin{equation}
\frac{4\pi m\rho _{v}^{2}}{\left( l^{2}+\rho _{v}^{2}\right) ^{3/2}}%
=2L_{v}-n_{av}+1\text{.}  \eqnum{16}
\end{equation}
From this expression, the threshold value of the magnetic moment (when the
potential barrier appears) is obtained as 
\begin{equation}
m^{\ast }=\left( \frac{3}{4}\right) ^{3/2}\frac{\left(
2L_{v}-n_{av}+1\right) l}{\pi }\text{.}  \eqnum{17}  \label{prag}
\end{equation}
For lower values of the magnetic moment, the energetic barrier between
vortices disappears and annihilation can not be prevented. Analogously, for
fixed magnetic moment, increasing vorticity $L_{v}$ of the vortex will make
the attraction stronger and for a certain value of $L_{v}$, antivortices
will be able to overcome the barrier. This critical value of vorticity can
also be estimated from Eq.~(\ref{prag}). One should notice that we leave the
possibility of $L_{v}\neq n_{av}$, which corresponds to the experimental
situation when first positive external flux lines are pinned by the magnetic
center, and then the polarity of the applied field is changed. Using $\Phi
^{+}/\Phi _{0}=4\sqrt{3}\pi /9~m/l$ for out-of-plane magnetic dipole
polarization, we find for $m=m^{\ast }$ that the critical condition becomes $%
\Phi ^{+}/\Phi _{0}=\left( 2L_{v}-n_{av}+1\right) /2$, where $\Phi ^{+}$
denotes the flux through the region of positive stray field of the dipole.
One should notice that the first stable vortex-antivortex pair ($%
L_{v}=n_{av}=1$) appears for $\Phi ^{+}=\Phi _{0}$, and further increase of
vorticity is a quantized process, with $\Delta \Phi ^{+}=\Phi _{0}/2$
necessary for stability for each one-unit-increase of $L_{v}$ and $n_{av}$.
Also in this case we cannot define a path around the positive stray field
region where the superconducting current is zero (see Fig.~\ref{fig5}(a))
and consequently the flux quantization condition does not apply here.

In this treatment, we assumed the presence of a giant vortex under the
dipole as an energetically preferable state. It is well known that in
infinite superconductors, thus in the absence of boundaries imposing the
symmetry of the superconducting state, the giant vortex splits into
multivortices. Since in our case an inhomogeneous applied magnetic field
dictates the behavior of superconducting electrons, it is not clear which
state carries less energy. Which central vortex configuration is realized
depends on the parameters of the superconductor, i.e. $\xi $, $\lambda $ and 
$d$, which come into play through the self energy of the involved vortices.
Therefore, for a particular superconducting film, we extend our approach to
the case of multivortices surrounded by an multi-antivortex ring. Namely, we
investigate the stability of $N$ vortex-antivortex pairs symmetrically
arranged around the dipole, where vortices sit on a ring with radius $\rho
_{v}$, and antivortices occupy the corresponding positions on the ring with
radius $\rho _{av}$. We apply the same approach as before, calculating the
interaction energy in a similar manner as in Eq.~(\ref{inten}), where the
interaction between each two vortices is included. After minimization of the
interaction energy with respect to the parameters $\rho _{v}$ and $\rho
_{av} $, we obtain numerically the energetically favorable positions of the
vortices. The results are shown in Fig.~\ref{fig10}, for the case of $N=3$
(see inset of Fig.~\ref{fig10}). Increasing magnetization of the dipole
increases the distance between vortices and antivortices in a way that
vortices come closer to the dipole and antivortices are repelled further
away. It should be noted that a certain critical value of the magnetization
is needed to prevent annihilation. In Fig.~\ref{fig11} we give this
threshold value as function of $l$ for different values of $N$. One should
note that this value of magnetic moment implies again the quantization of
the penetrating flux (as in the case of a giant vortex surrounded by
antivortices), but with $\Delta \Phi ^{+}=1.0489\Phi _{0}$ necessary for
stability of the first pair, and slightly decreasing for additional pairs.
For $N>4$, $\Delta \Phi ^{+}$ becomes smaller than the flux quantum. The
interpretation of this behavior can be that multivortices eventually join
into a giant vortex, or that a different geometry of the superconducting
state appears. For example, for larger number of pairs, due to the large
linear density of vortices along the ring, they could rearrange, forming
more than a single ring.

Fig.~\ref{fig11} gives only the critical condition for stability of
vortex-antivortex pairs. However, which configuration has the lowest energy
and is thus the energetically preferable state cannot be inferred from this
figure. In order to compare the energies of states with different number of
pairs, i.e. $N$, we include the self-energy of individual vortices in the
calculation 
\begin{equation}
F_{v}=\frac{\Phi _{0}^{2}}{16\pi ^{2}\lambda }\left[ d~ln\frac{\lambda }{\xi 
}+2tanh\left( \frac{d}{2}\right) ~ln\left( 1+cotanh\left( \frac{d}{2}\right)
\right) \right] .  \eqnum{18}
\end{equation}
The numerical results, as referred to the Meissner state, are shown in Fig.~%
\ref{fig12}, for a thin SC film with $\lambda /\xi =10$. With increasing
magnetic moment of the dipole the energetically favorable state goes through
successive states in which $N$ increases. If we calculate again the flux $%
\Delta \Phi ^{+}$ which now corresponds to the appearance of the next $N$
state as the ground state, we find that $\Delta \Phi ^{+}=1.97\Phi _{0}$,
for the appearance of the first pair, and for $N\geq 2$, the additional flux
slightly decreases with increasing $N$, starting from $\Delta \Phi
^{+}=1.089\Phi _{0}$, for $N=2$. Larger flux needed for the first
vortex-antivortex state can be explained by the fact that this cylindrically
asymmetrical state appears in a symmetrical magnetic potential. For $N\geq 2$%
, the results correspond to our previous analysis. It should be stressed
that these results depend on the parameters of the superconductor, namely $%
\lambda $ and $\xi $, which come into play through the self-energy of the
vortices. This is in contrast to the interaction energy (see the inset of
Fig.~\ref{fig12}) which, in the London approximation, is independent of $\xi 
$.

\section{Conclusion}

To summarize, we applied the London theory to investigate flux pinning in SC
films due to the presence of a magnetic dipole situated above (or under) the
SC. Depending on the direction of the dipole magnetic moment, we obtained
exact analytic expressions for the MD-vortex interaction energy and
screening currents. We obtained the asymptotic behavior of the interaction
potential and the induced currents for specific values of the involved
parameters. We calculated the pinning potential for both an in- and
out-of-plane magnetized dipole. Our results show that an out-of-plane
magnetized dipole attracts a vortex if aligned parallel to it, and opposite,
for anti-parallel alignment the vortex is repelled. This is a consequence of
the mono-directional superconducting current induced in the SC for
out-of-plane magnetization of the MD. However, for in-plane magnetization,
the dipole-vortex interaction shows a dual behavior, namely, the vortex is
attracted to the negative pole of the MD and repelled on the other side.
Moreover, the position of the pinning site depends on the position of the MD
and thickness of the SC. We calculated these dependences and showed that it
is linear for thin superconductors, or large MD-vortex distances. Due to the
dual behavior of the pinning potential, we explored the possible
co-existence of vortices and antivortices in such systems. The total
interaction energy calculation leads to the conclusion that the vortex and
antivortex are separated by an energy barrier due to the short range
interaction with the dipole, and therefore, these pairs could be stable.
Both in- and out-of-plane magnetized dipoles are able to keep these vortices
apart. We calculated analytically the interaction potential in the presence
of vortex-antivortex pairs (or giant vortex-single antivortices) and gave
estimates of the parameters necessary for stability of such fascinating
configurations in a thin SC film.\bigskip

\section*{ACKNOWLEDGMENTS}

\bigskip

This work was supported by the Flemish Science Foundation (FWO-Vl), the
Belgian Inter-University Attraction Poles (IUAP), the ``Onderzoeksraad van
de Universiteit Antwerpen'' (GOA), and the ESF programme on ``Vortex
matter''. Stimulating discussions with D.~Vodolazov, V.~V.~Moshchalkov, and
M.~Van~Bael are gratefully acknowledged.

\bigskip

\section*{APPENDIX: The magnetic field of a magnetic dipole in the presence
of a SC film}

\bigskip

The magnetic field of a vortex $\vec{h}_{v}=rot\vec{A}_{v}$, which is
perpendicular to the plane of the film, is determined as the solution of the
system of equations 
\begin{equation}
\vec{A}_{v}\left( \vec{\rho},z\right) +\lambda ^{2}%
\mathop{\rm rot}%
\mathop{\rm rot}%
\vec{A}_{v}\left( \vec{\rho},z\right) =\vec{\Phi}\left( \vec{\rho}-\vec{\rho}%
_{v}\right) ,\qquad \left| z\right| <d/2,  \eqnum{A1a}  \label{VPvort1}
\end{equation}
\begin{equation}
-\nabla ^{2}\vec{A}_{v}\left( \vec{\rho},z\right) =0\qquad \left| z\right|
>d/2,  \eqnum{A1b}  \label{VPvort2}
\end{equation}
with the following boundary conditions: i) the continuity of the vector
potential components at $z=\pm d/2$,\ and ii) their vanishing far from the
superconductor (at $\left| z\right| \rightarrow \infty $). Here $\vec{\Phi}%
\left( \vec{\rho}-\vec{\rho}_{v}\right) $ takes into account the vortex,
where $\vec{\rho}_{v}$ is the vortex position in the film plane. In
cylindrical coordinates $\Phi _{\rho }=\Phi _{z}=0$, $\Phi _{\varphi }=L\Phi
_{0}/2\pi \left| \vec{\rho}-\vec{\rho}_{v}\right| $ (so that $%
\mathop{\rm rot}%
\vec{\Phi}=L\Phi _{0}\delta \left( \vec{\rho}-\vec{\rho}_{v}\right) $). The
solutions of Eqs.~(\ref{VPvort1})-(\ref{VPvort2}) are~\cite{vfield1,ita} 
\begin{equation}
A_{v\varphi }^{\left( i\right) }\left( \vec{\rho},z\right) =\frac{L\Phi _{0}%
}{2\pi \lambda }\int\limits_{0}^{\infty }dq\;\frac{J_{1}\left( qR/\lambda
\right) }{k^{2}}\left[ 1-\frac{kq\cosh \left( kz/\lambda \right) }{Q\sinh
\left( kd/2\lambda \right) }\right] ,  \eqnum{A2a}
\end{equation}
\begin{equation}
A_{v\varphi }^{\left( o\right) }\left( \vec{\rho},z\right) =\frac{L\Phi _{0}%
}{2\pi \lambda }\int\limits_{0}^{\infty }dq\frac{J_{1}\left( qR/\lambda
\right) }{Q}\exp \left( -q\frac{2\left| z\right| -d}{2\lambda }\right) . 
\eqnum{A2b}  \label{Av}
\end{equation}
Here $k=\left( 1+q^{2}\right) ^{1/2},\;Q=k\left[ k+q\coth \left( kd/2\lambda
\right) \right] ,\;R=\left| \vec{\rho}-\vec{\rho}_{v}\right| =\left[ \rho
^{2}+\rho _{v}^{2}-2\rho \rho _{v}\cos \left( \varphi -\varphi _{v}\right) %
\right] ^{1/2},$ $J_{\nu }\left( x\right) $ is the Bessel function, index ``$%
i$($o$)'' denotes the field inside (outside) the superconductor. The
components of the vortex magnetic field are given by~ \cite{ita} 
\begin{equation}
h_{vz}^{(i)}\left( \rho ,z\right) =\frac{L\Phi _{0}}{2\pi \lambda ^{2}}\left[
K_{0}\left( \frac{R}{\lambda }\right) -\int\limits_{0}^{\infty }\frac{%
dq\,q^{2}}{kQ}J_{0}\left( \frac{qR}{\lambda }\right) \frac{\cosh \left(
kz/\lambda \right) }{\sinh \left( kd/2\lambda \right) }\right] ,  \eqnum{A3a}
\end{equation}
\begin{equation}
h_{v\rho }^{(i)}\left( \rho ,z\right) =\frac{L\Phi _{0}}{2\pi \lambda ^{2}}%
\int\limits_{0}^{\infty }\frac{dq\,q}{Q}J_{1}\left( \frac{qR}{\lambda }%
\right) \frac{\sinh \left( kz/\lambda \right) }{\sinh \left( kd/2\lambda
\right) },  \eqnum{A3b}
\end{equation}
\begin{equation}
h_{vz}^{(o)}\left( \rho ,z\right) =\frac{L\Phi _{0}}{2\pi \lambda ^{2}}%
\int\limits_{0}^{\infty }\frac{dq\,q}{Q}J_{0}\left( \frac{qR}{\lambda }%
\right) \exp \left( -q\frac{2\left| z\right| -d}{2\lambda }\right) , 
\eqnum{A3c}
\end{equation}
\begin{equation}
h_{v\rho }^{(o)}\left( \rho ,z\right) =\frac{L\Phi _{0}}{2\pi \lambda ^{2}}%
\mathop{\rm sgn}%
(z)\int\limits_{0}^{\infty }\frac{dq\,q}{Q}J_{1}\left( \frac{qR}{\lambda }%
\right) \exp \left( -q\frac{2\left| z\right| -d}{2\lambda }\right) , 
\eqnum{A3d}  \label{Hvr2}
\end{equation}
\begin{equation}
h_{v\varphi }^{(i)}=h_{v\varphi }^{(o)}=0,  \eqnum{A3e}
\end{equation}
where $K_{0}\left( x\right) $ is the MacDonald function.

The magnetic field of a dipole, situated at the distance $l$\ above the
film, can be written in the following way (here and below $l=\left| a\right|
-d/2$) (see, for example, Ref.~\cite{badia}) 
\begin{equation}
\vec{h}_{m}\left( \vec{\rho},z\right) =\left\{ 
\begin{array}{l}
\vec{h}_{d}\left( \vec{\rho},z\right) +\vec{h}_{m,+}^{(o)}\left( \vec{\rho}%
,z\right) ,~~~~z>d/2, \\ 
\vec{h}_{m}^{(i)}\left( \vec{\rho},z\right) ,\qquad \qquad \qquad -d/2\leq
z\leq d/2, \\ 
\vec{h}_{m,-}^{(o)}\left( \vec{\rho},z\right) ,\qquad \qquad \qquad z<-d/2,
\end{array}
\right.  \eqnum{A4}  \label{Hdot}
\end{equation}
where $\vec{h}_{d}\left( \vec{\rho},z\right) $ is the direct contribution
from the dipole, $\vec{h}_{m,+}^{(o)}\left( \vec{\rho},z\right) $ is the
induced field above the superconductor, $\vec{h}_{m}^{(i)}\left( \vec{\rho}%
\right) $ is the dipole field penetrating inside the superconductor and $%
\vec{h}_{m,-}^{(o)}\left( \vec{\rho}\right) $ denotes the field of the
dipole below the film. The vector potential $\vec{A}_{d}\left( \vec{\rho}%
,z\right) $ is defined by the equation 
\begin{equation}
\mathop{\rm rot}%
\vec{A}_{d}\left( \vec{\rho},z\right) =4\pi \vec{m}\delta \left( \vec{\rho}%
\right) \delta \left( z-a\right) ,  \eqnum{A5}  \label{VPdipole}
\end{equation}
where $\vec{m}$ is the magnetic moment of the dipole. The potential $\vec{A}%
_{m}^{(i)}\left( \vec{\rho}\right) $\ is found as the solution of Eq.~(\ref
{VPvort1}) with zero right side and $\vec{A}_{m,\pm }^{(o)}\left( \vec{\rho}%
\right) $ as solution of Eq.~(\ref{VPvort2}). The integration constants have
to be found through similar boundary conditions as for the vortex field
distribution: i) the continuity of the vector potential components at $z=\pm
d/2$\ and ii) their vanishing at $\left| z\right| \rightarrow \infty $.

We consider two orientations of the MD magnetization.

I) The dipole is directed {\it perpendicular} to the film plane, i.e. $%
m_{\rho }=m_{\varphi }=0,\;m_{z}=m$. In this case the vector potential has
only an azimuthal component and is described by the following expressions: 
\begin{equation}
A_{d\varphi }(\rho ,z)=\frac{m}{\lambda ^{2}}\int_{0}^{\infty }dq\
qJ_{1}\left( \frac{q\rho }{\lambda }\right) \exp \left( -\frac{q}{\lambda }%
\left| z-a\right| \right) =\frac{m\rho }{\left[ \rho ^{2}+\left( z-a\right)
^{2}\right] ^{3/2}},  \eqnum{A6a}
\end{equation}
\begin{equation}
A_{m\varphi }^{\left( i\right) }(\rho ,z)=\frac{1}{\lambda ^{2}}%
\int_{0}^{\infty }dq\ qJ_{1}\left( \frac{q\rho }{\lambda }\right) \left[
B_{1}\left( q\right) \exp \left( -\frac{kz}{\lambda }\right) +D_{1}\left(
q\right) \exp \left( \frac{kz}{\lambda }\right) \right] ,  \eqnum{A6b}
\end{equation}
\begin{equation}
A_{m\varphi ,\pm }^{\left( o\right) }(\rho ,z)=\frac{1}{\lambda ^{2}}%
\int_{0}^{\infty }dq\ qJ_{1}\left( \frac{q\rho }{\lambda }\right) B_{2,\pm
}\left( q\right) \exp \left( -\frac{qz}{\lambda }\right) ,  \eqnum{A6c}
\end{equation}
where $B_{1}\left( q\right) =B(q)q\left( k-q\right) \exp \left( -kd/2\lambda
\right) ,~D_{1}\left( q\right) =B(q)q\left( k+q\right) \exp \left(
kd/2\lambda \right) ,~B_{2,+}\left( q\right) =-B(q)\sinh \left( kd/\lambda
\right) \exp \left( qd/2\lambda \right) ,~$and$~B_{2,-}\left( q\right)
=2B(q)kq\exp \left( qd/2\lambda \right) ,~$with$~B\left( q\right) =2m\exp
\left( -ql/\lambda \right) /\left( \left( k+q\right) ^{2}\exp \left(
kd/\lambda \right) -\left( k-q\right) ^{2}\exp \left( -kd/\lambda \right)
\right) $.

Consequently, we obtain the following components of the dipole magnetic
field 
\begin{equation}
h_{mz}^{(i)}\left( \rho ,z\right) =\frac{1}{\lambda ^{3}}\int_{0}^{\infty
}dq\ q^{2}J_{0}\left( \frac{q\rho }{\lambda }\right) \left[ B_{1}\left(
q\right) \exp \left( -\frac{kz}{\lambda }\right) +D_{1}\left( q\right) \exp
\left( \frac{kz}{\lambda }\right) \right] ,  \eqnum{A7a}
\end{equation}
\begin{equation}
h_{m\rho }^{(i)}\left( \rho ,z\right) =\frac{1}{\lambda ^{3}}%
\int_{0}^{\infty }dq\ kqJ_{1}\left( \frac{q\rho }{\lambda }\right) \left[
B_{1}\left( q\right) \exp \left( -\frac{kz}{\lambda }\right) -D_{1}\left(
q\right) \exp \left( \frac{kz}{\lambda }\right) \right] ,  \eqnum{A7b}
\end{equation}
\begin{equation}
h_{dz}\left( \rho ,z\right) =\frac{m}{\lambda ^{3}}\int_{0}^{\infty }dq\
q^{2}J_{0}\left( \frac{q\rho }{\lambda }\right) \exp \left( -\frac{q}{%
\lambda }\left| z-a\right| \right) =\frac{m\left[ 2\left( z-a\right)
^{2}-\rho ^{2}\right] }{\left[ \rho ^{2}+\left( z-a\right) ^{2}\right] ^{5/2}%
},  \eqnum{A7c}
\end{equation}
\begin{equation}
h_{d\rho }\left( \rho ,z\right) =\frac{m%
\mathop{\rm sgn}%
\left( z-a\right) }{\lambda ^{3}}\int_{0}^{\infty }dq\ q^{2}J_{1}\left( 
\frac{q\rho }{\lambda }\right) \exp \left( -\frac{q}{\lambda }\left|
z-a\right| \right) =\frac{3m\rho \left( z-a\right) }{\left[ \rho ^{2}+\left(
z-a\right) ^{2}\right] ^{5/2}},  \eqnum{A7d}
\end{equation}
\begin{equation}
h_{mz,\pm }^{(o)}\left( \rho ,z\right) =\frac{1}{\lambda ^{3}}%
\int_{0}^{\infty }dq\ q^{2}J_{0}\left( \frac{q\rho }{\lambda }\right)
B_{2,\pm }\left( q\right) \exp \left( -\frac{qz}{\lambda }\right) , 
\eqnum{A7e}
\end{equation}
\begin{equation}
h_{m\rho ,\pm }^{(o)}\left( \rho ,z\right) =\frac{1}{\lambda ^{3}}%
\int_{0}^{\infty }dq\ q^{2}J_{1}\left( \frac{q\rho }{\lambda }\right)
B_{2,\pm }\left( q\right) \exp \left( -\frac{qz}{\lambda }\right) , 
\eqnum{A7f}
\end{equation}
\begin{equation}
h_{m\varphi }^{(1)}=h_{m\varphi ,\pm }^{(2)}=0.  \eqnum{A7g}
\end{equation}

II) The dipole is magnetized in the direction {\it parallel} to the SC film
plane (in-plane magnetization). Following the same procedure as above, we
obtain 
\begin{equation}
h_{d\rho }\left( \rho ,\varphi ,z\right) =\frac{m\cos \varphi }{2\lambda ^{3}%
}\int_{0}^{\infty }dq\ q^{2}\exp \left( -\frac{q}{\lambda }\left| z-a\right|
\right) \left[ J_{2}\left( \frac{q\rho }{\lambda }\right) -J_{0}\left( \frac{%
q\rho }{\lambda }\right) \right] =\frac{m\left[ 2\rho ^{2}-\left( z-a\right)
^{2}\right] \cos \varphi }{\left[ \rho ^{2}+\left( z-a\right) ^{2}\right]
^{5/2}},  \eqnum{A8a}
\end{equation}
\begin{equation}
h_{d\varphi }\left( \rho ,\varphi ,z\right) =\frac{m\sin \varphi }{\lambda
^{2}\rho }\int_{0}^{\infty }dq\ q\exp \left( -\frac{q}{\lambda }\left|
z-a\right| \right) J_{1}\left( \frac{q\rho }{\lambda }\right) =\frac{m\sin
\varphi }{\left[ \rho ^{2}+\left( z-a\right) ^{2}\right] ^{3/2}}, 
\eqnum{A8b}
\end{equation}
\begin{equation}
h_{dz}\left( \rho ,\varphi ,z\right) =-\frac{m\cos \varphi }{\lambda ^{3}}%
\int_{0}^{\infty }dq\ q^{2}\exp \left( -\frac{q}{\lambda }\left| z-a\right|
\right) J_{1}\left( \frac{q\rho }{\lambda }\right) =-\frac{3m\rho \left|
z-a\right| \cos \varphi }{\left[ \rho ^{2}+\left( z-a\right) ^{2}\right]
^{5/2}},  \eqnum{A8c}
\end{equation}
\begin{eqnarray}
h_{m\rho }^{(i)}\left( \rho ,\varphi ,z\right) &=&\frac{\cos \varphi }{%
2\lambda ^{3}}\int_{0}^{\infty }dq\ kq\left[ J_{2}\left( \frac{q\rho }{%
\lambda }\right) -J_{0}\left( \frac{q\rho }{\lambda }\right) \right] 
\nonumber \\
&&\times \left[ D_{1}\left( q\right) \exp \left( \frac{kz}{\lambda }\right)
-B_{1}\left( q\right) \exp \left( -\frac{kz}{\lambda }\right) \right] , 
\eqnum{A8d}
\end{eqnarray}
\begin{equation}
h_{m\varphi }^{(i)}\left( \rho ,\varphi ,z\right) =\frac{\sin \varphi }{%
\lambda ^{2}\rho }\int_{0}^{\infty }dq\ kJ_{1}\left( \frac{q\rho }{\lambda }%
\right) \left[ D_{1}\left( q\right) \exp \left( \frac{kz}{\lambda }\right)
-B_{1}\left( q\right) \exp \left( -\frac{kz}{\lambda }\right) \right] , 
\eqnum{A8e}
\end{equation}
\begin{equation}
h_{mz}^{(i)}\left( \rho ,\varphi ,z\right) =-\frac{\cos \varphi }{\lambda
^{3}}\int_{0}^{\infty }dq\ q^{2}J_{1}\left( \frac{q\rho }{\lambda }\right) %
\left[ D_{1}\left( q\right) \exp \left( \frac{kz}{\lambda }\right)
+B_{1}\left( q\right) \exp \left( -\frac{kz}{\lambda }\right) \right] , 
\eqnum{A8f}
\end{equation}
\begin{equation}
h_{m\rho ,\pm }^{(o)}\left( \rho ,\varphi ,z\right) =-\frac{\cos \varphi }{%
2\lambda ^{3}}\int_{0}^{\infty }dq\ q^{2}\exp \left( -\frac{qz}{\lambda }%
\right) \left[ J_{2}\left( \frac{q\rho }{\lambda }\right) -J_{0}\left( \frac{%
q\rho }{\lambda }\right) \right] B_{2,\pm }\left( q\right) ,  \eqnum{A8g}
\end{equation}
\begin{equation}
h_{m\varphi ,\pm }^{(o)}\left( \rho ,\varphi ,z\right) =-\frac{\sin \varphi 
}{\lambda ^{2}\rho }\int_{0}^{\infty }dq\ q\exp \left( -\frac{qz}{\lambda }%
\right) J_{1}\left( \frac{q\rho }{\lambda }\right) B_{2,\pm }\left( q\right)
,  \eqnum{A8h}
\end{equation}
\begin{equation}
h_{mz,\pm }^{(o)}\left( \rho ,\varphi ,z\right) =-\frac{\cos \varphi }{%
\lambda ^{3}}\int_{0}^{\infty }dq\ q^{2}\exp \left( -\frac{qz}{\lambda }%
\right) J_{1}\left( \frac{q\rho }{\lambda }\right) B_{2,\pm }\left( q\right)
,  \eqnum{A8i}
\end{equation}

\begin{figure}[tbp]
\caption{An oblique view of the systems under investigation with (a) out-
and (b) in-plane directed magnetic dipole above a superconducting film
interacting with a vortex.}
\label{fig1}
\end{figure}
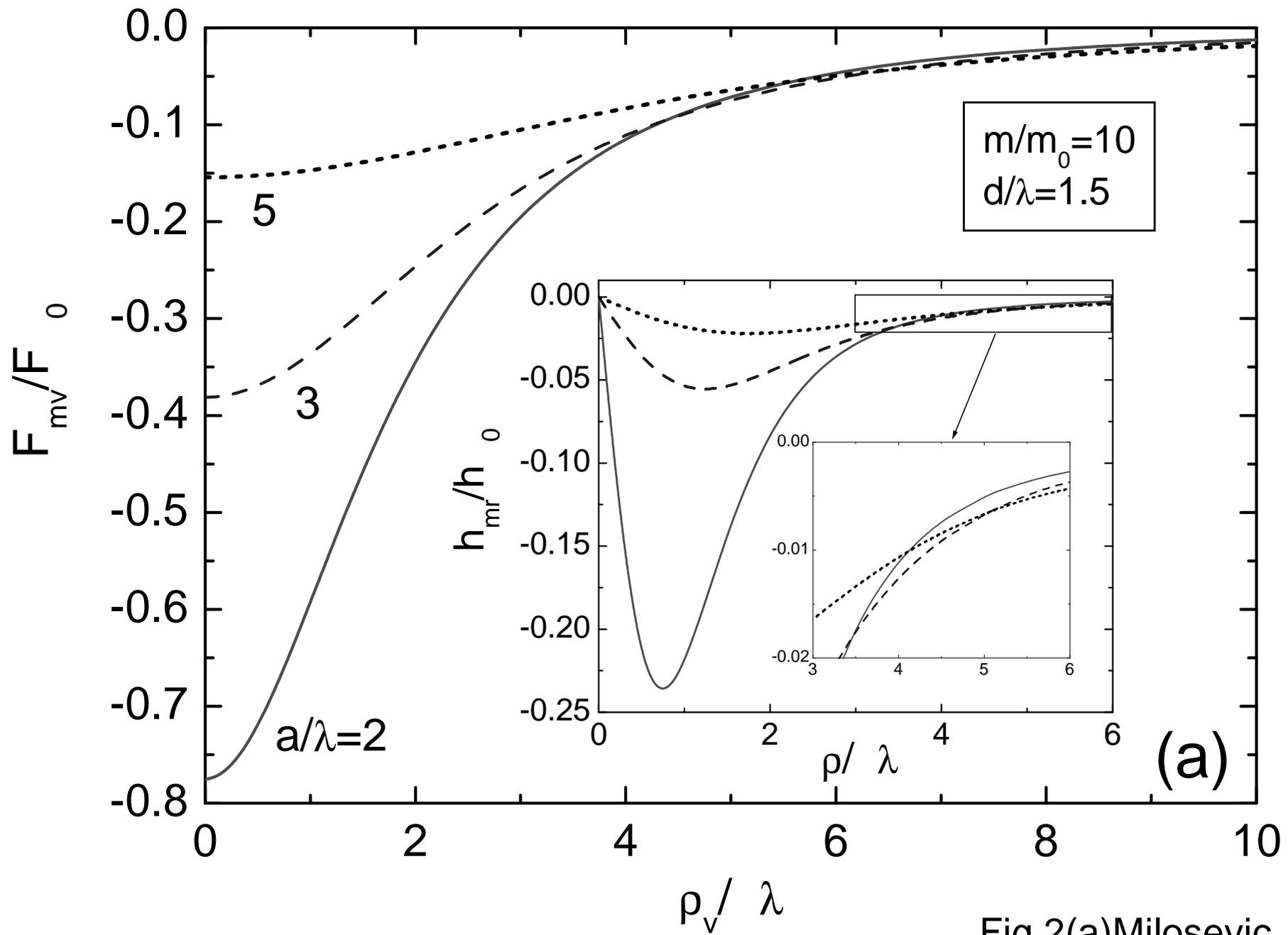
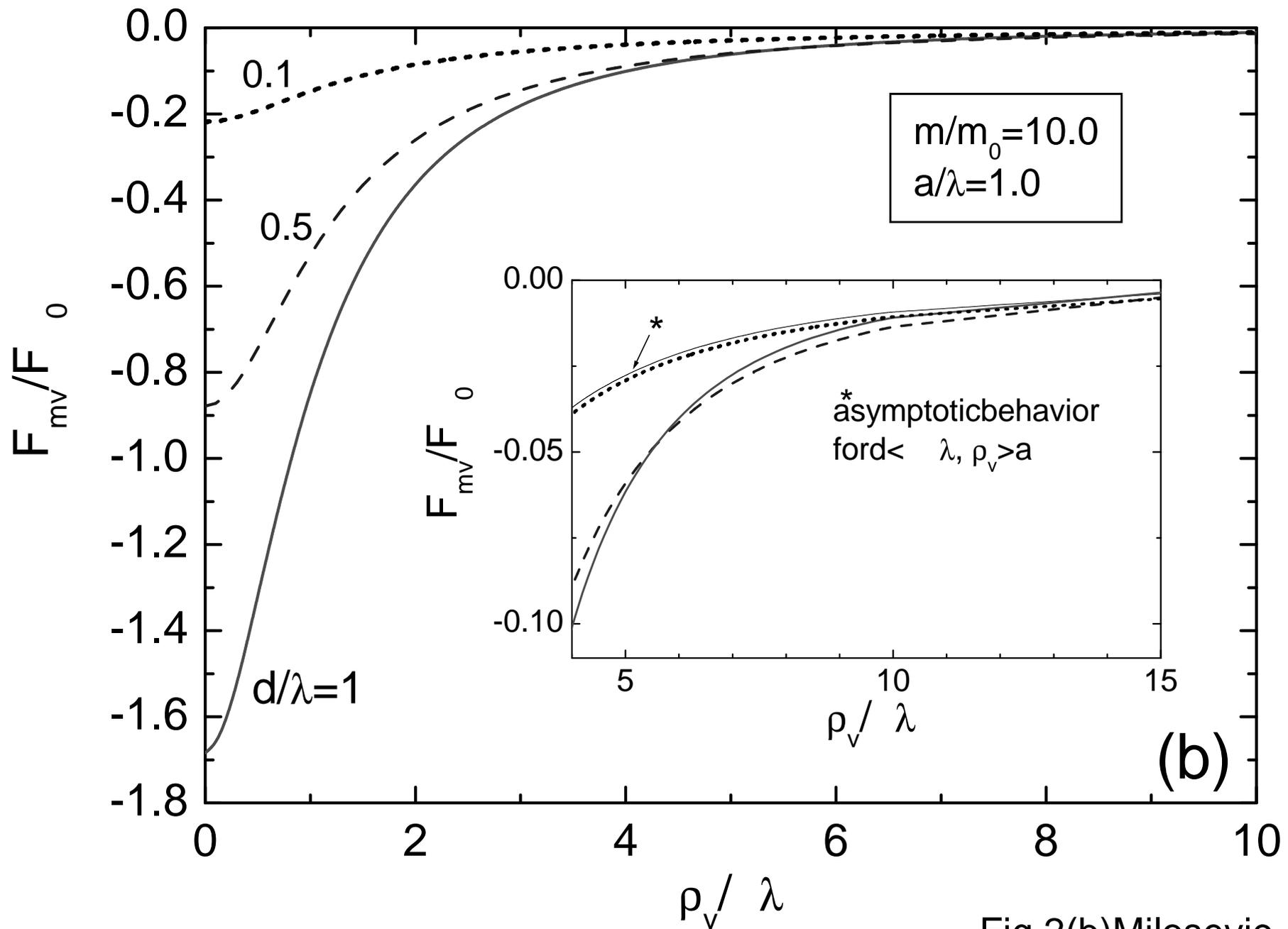
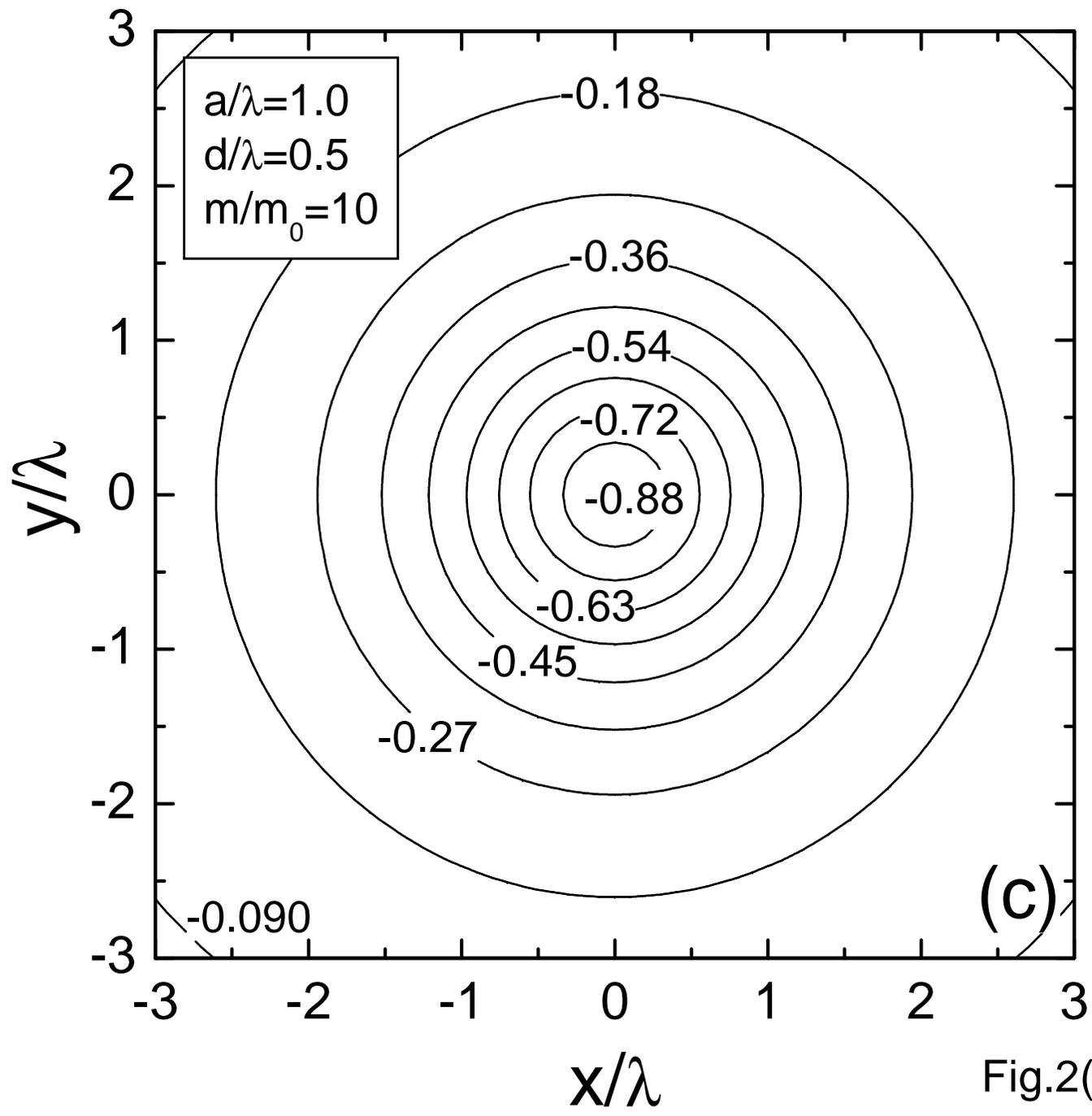
\begin{figure}[tbp]
\caption{The interaction energy of the vortex with an out-of-plane magnetic
dipole (MD) in units of $F_{0}=\Phi _{0}^{2}/(\protect\pi \protect\lambda )$
as a function of the position of the vortex ($\protect\rho _{v}$): a) for
different position of the MD; b) for different thickness of the
superconductor; and c) the contour plot of the interaction potential. Inset
in (a) illustrates the component of the total MD magnetic field in the
presence of the SC, tangential to the SC surface ($h_{0}=m_{0}/\protect%
\lambda ^{3}$). Inset in (b) depicts the asymptotic behavior for specific
values of the parameters. }
\label{fig2}
\end{figure}
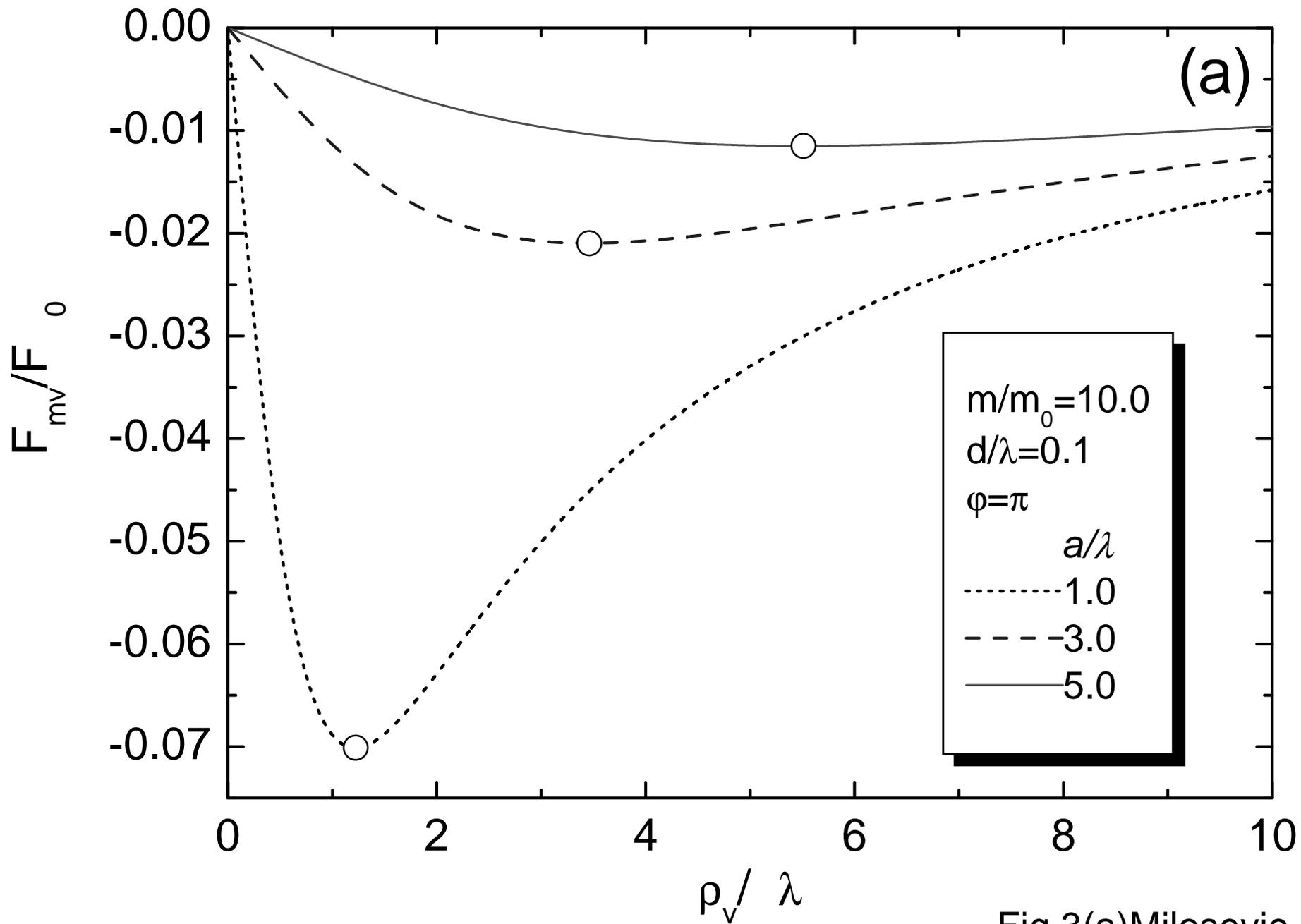
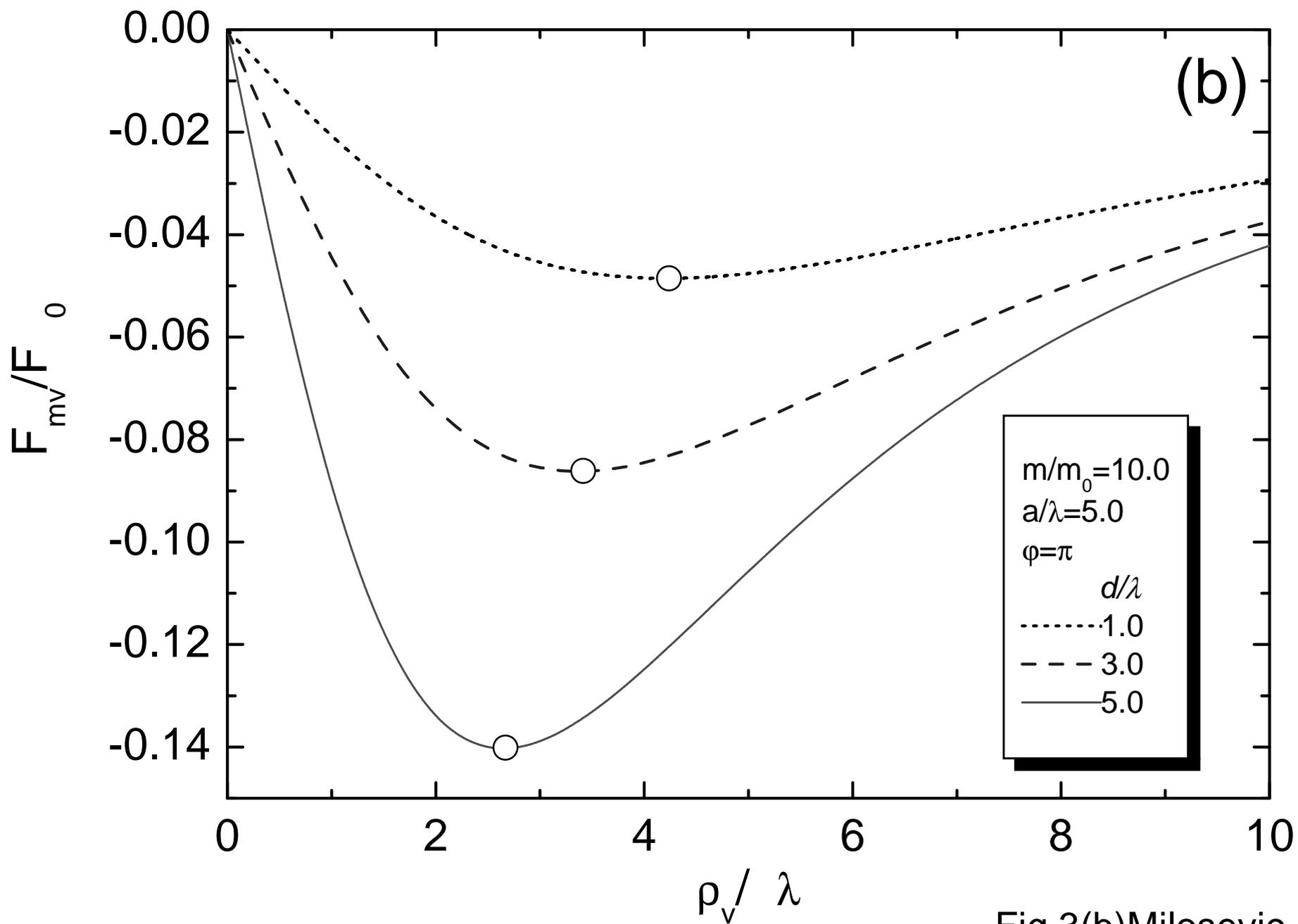
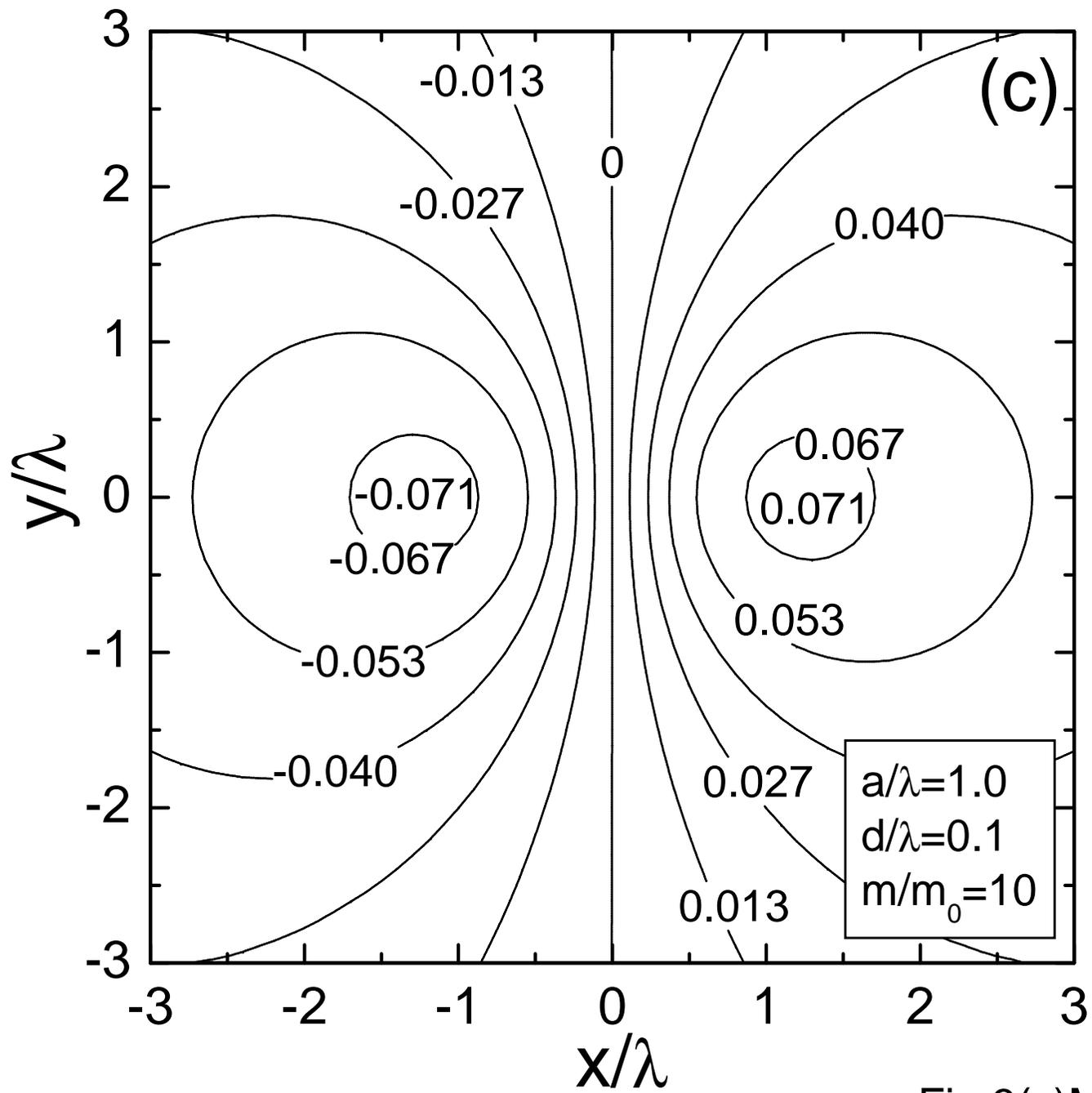
\begin{figure}[tbp]
\caption{The interaction energy of the vortex with an in-plane magnetic
dipole (MD) in units of $F_{0}=\Phi _{0}^{2}/(\protect\pi \protect\lambda )$
as a function of the position of the vortex ($\protect\rho _{v}$): a) for
different vertical position of the MD; b) for different thickness of the
superconductor; and c) the contour plot of the interaction potential. Open
circles denote the energetically favorable position of the vortex.}
\label{fig3}
\end{figure}
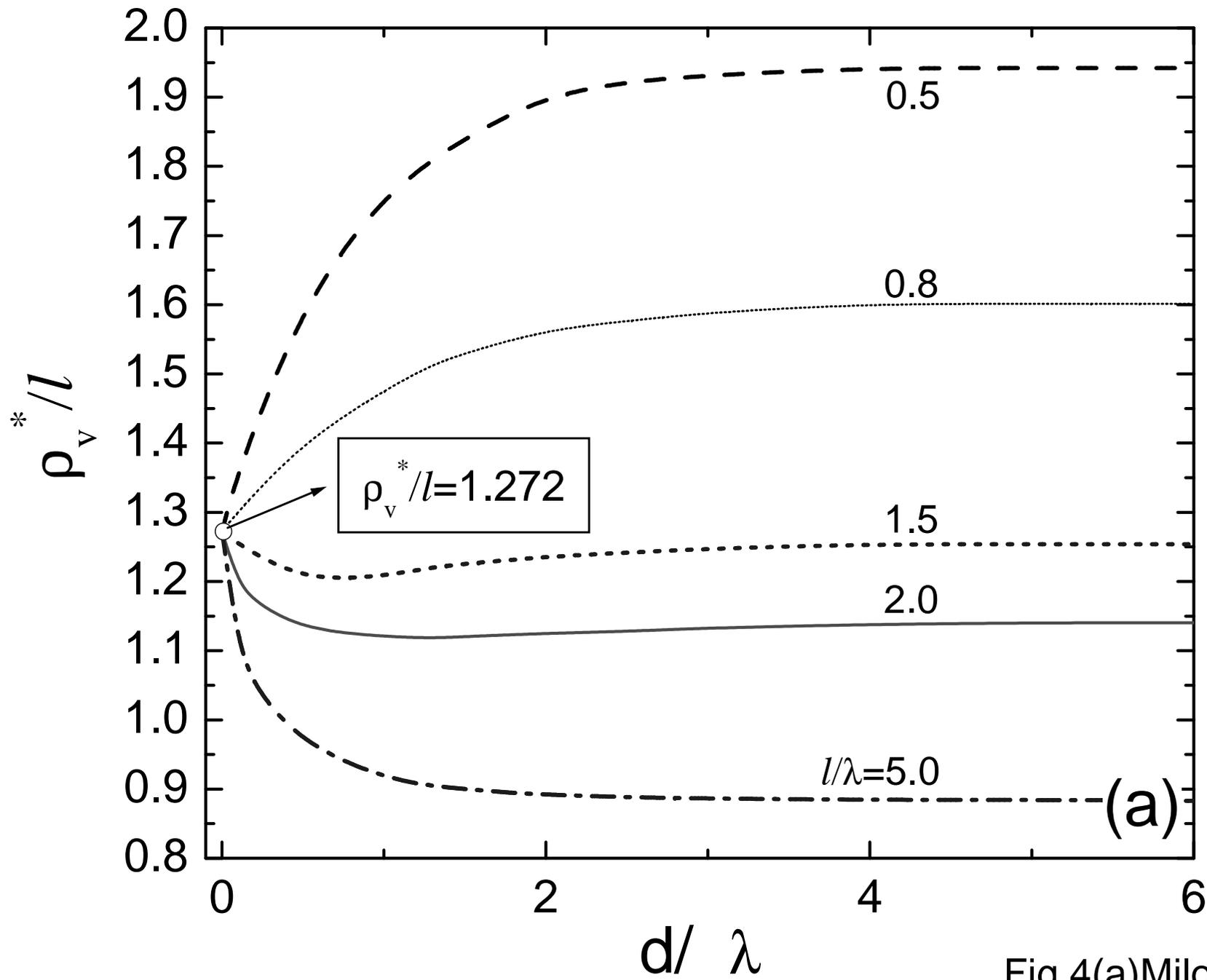
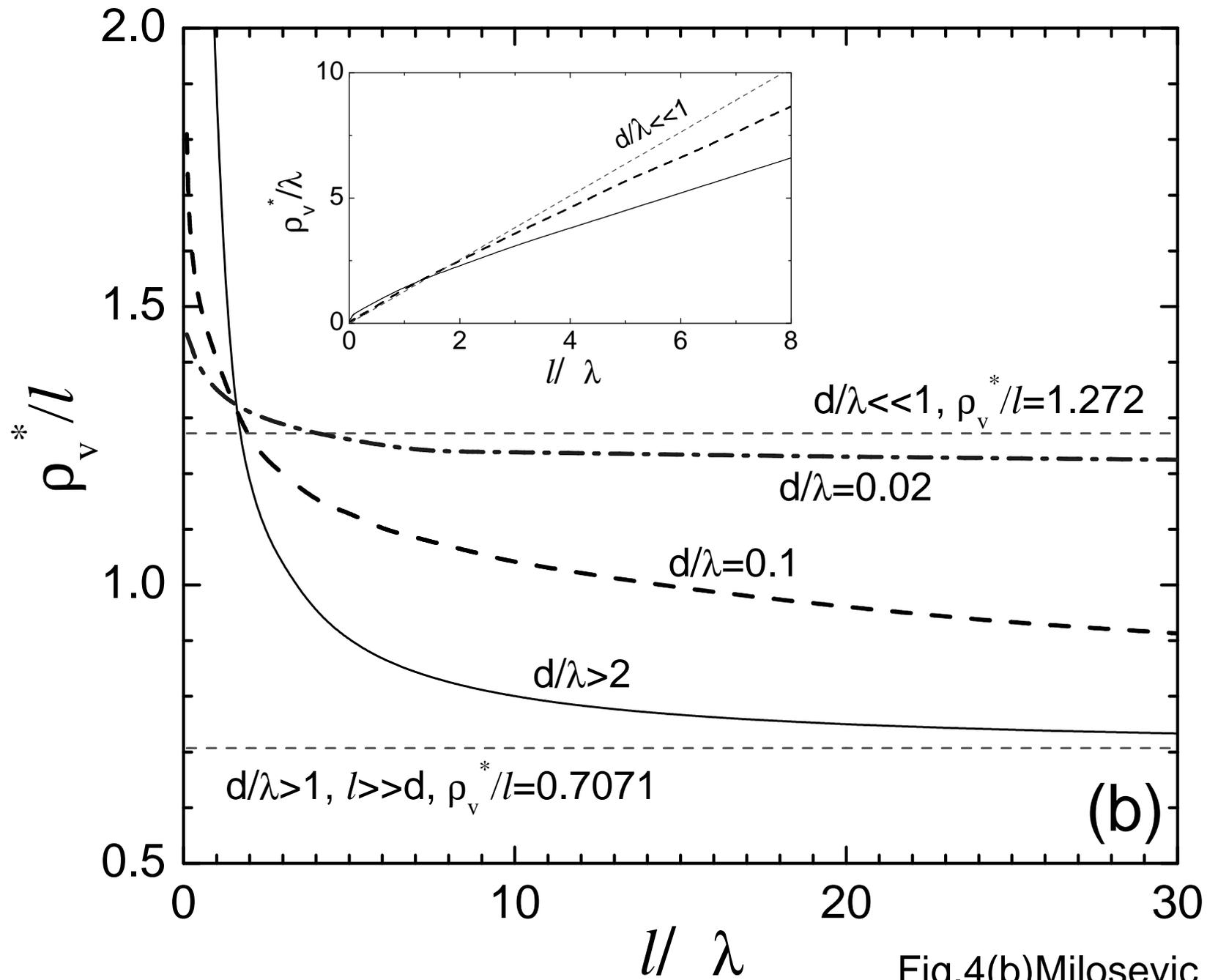
\begin{figure}[tbp]
\caption{The position of the vortex ($\protect\rho _{v}^{\ast }$) as
function of the thickness of the superconductor ($d$) and distance between
the in-plane magnetized dipole and the SC surface ($l=a-d/2$). $\protect\rho %
_{v}^{\ast }$ is scaled by $l$ or by $\protect\lambda $ (inset of (b)).}
\label{fig4}
\end{figure}
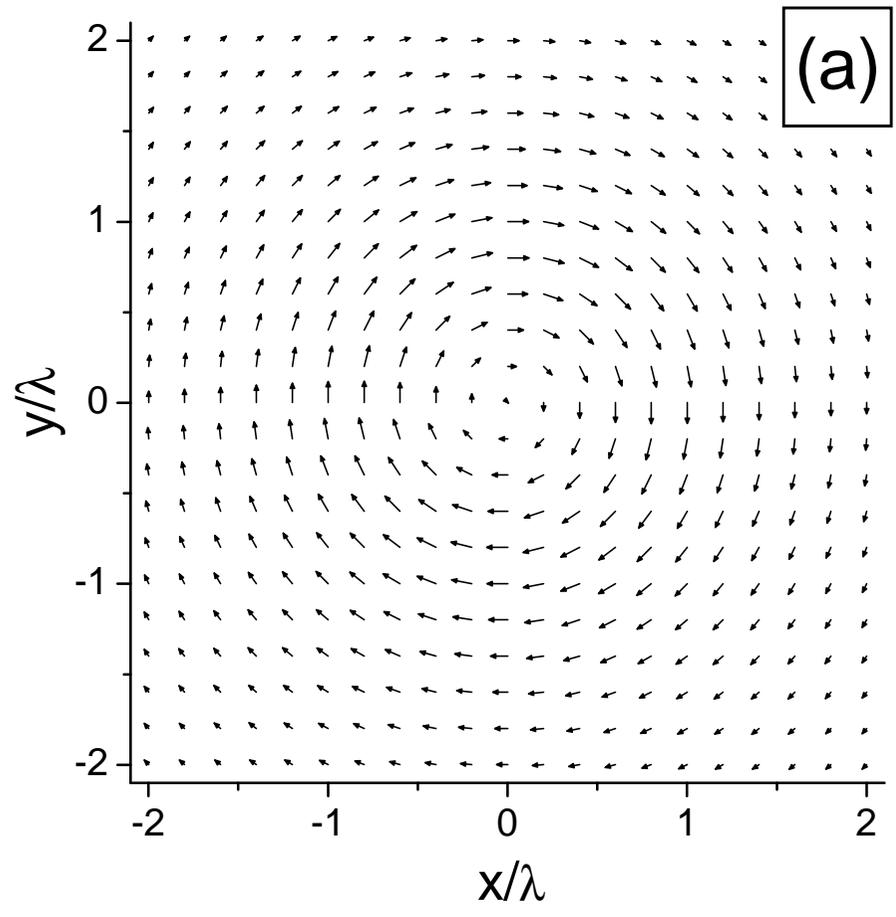
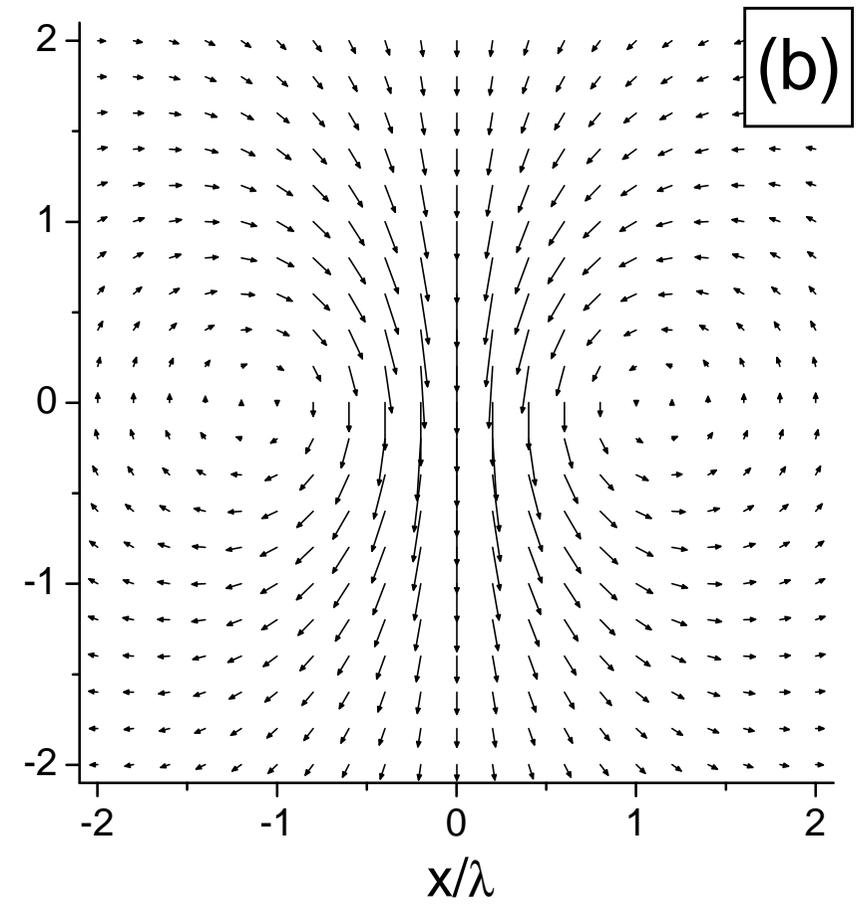
\begin{figure}[tbp]
\caption{The screening current density in the SC film with $d=0.5\protect%
\lambda $ for: a) out-of-plane and b) in-plane magnetized magnetic dipole
with $m=m_{0}$ positioned at $a=1.0\protect\lambda $ above the
superconducting film.}
\label{fig5}
\end{figure}
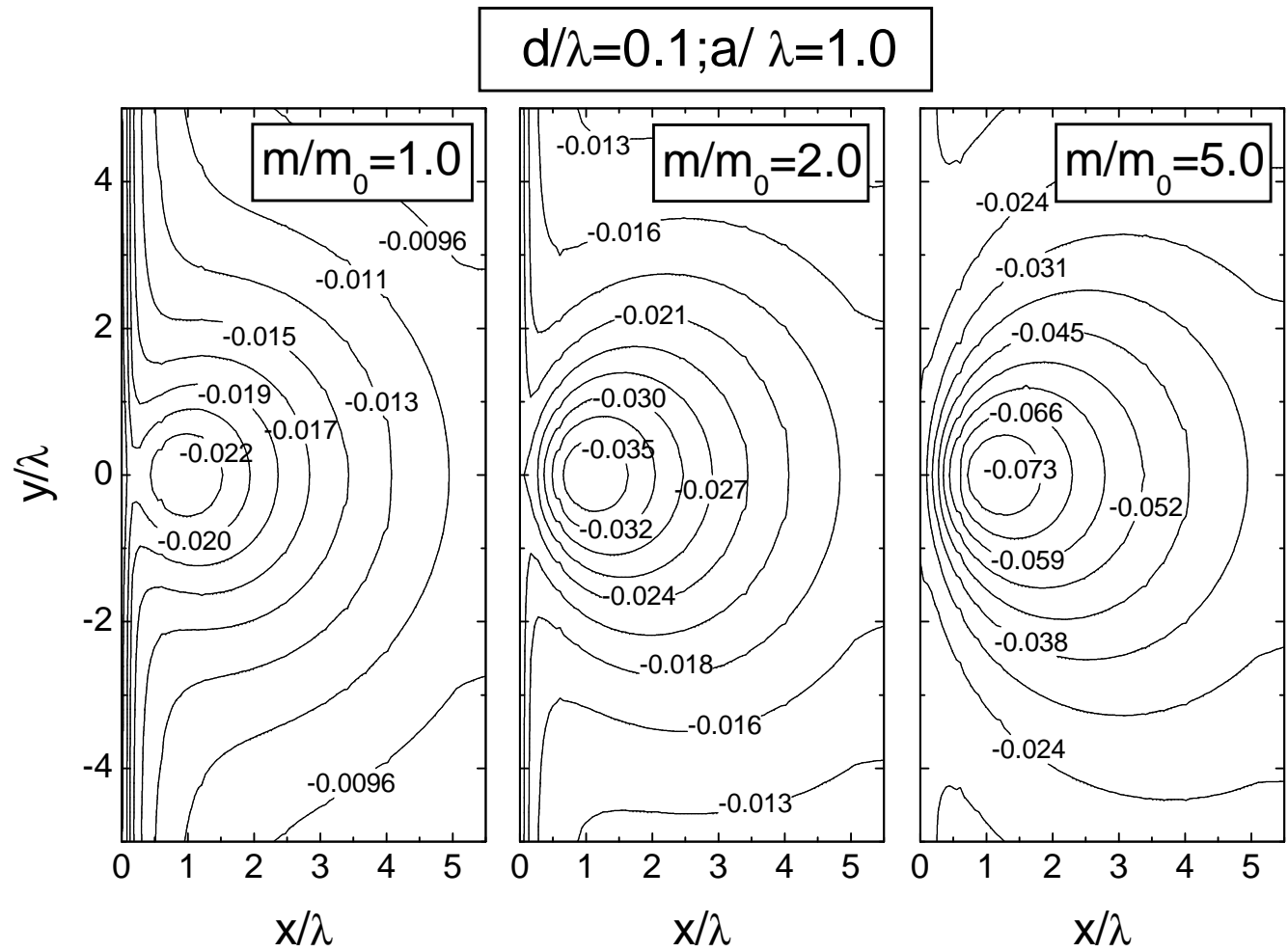
\begin{figure}[tbp]
\caption{The total interaction energy (in units of $F_{0}=\Phi _{0}^{2}/(%
\protect\pi \protect\lambda )$) between the in-plane magnetic dipole (MD)
and the vortex and the antivortex placed at opposite sides of the dipole for
different values of the magnetic moment $m$.}
\label{fig6}
\end{figure}
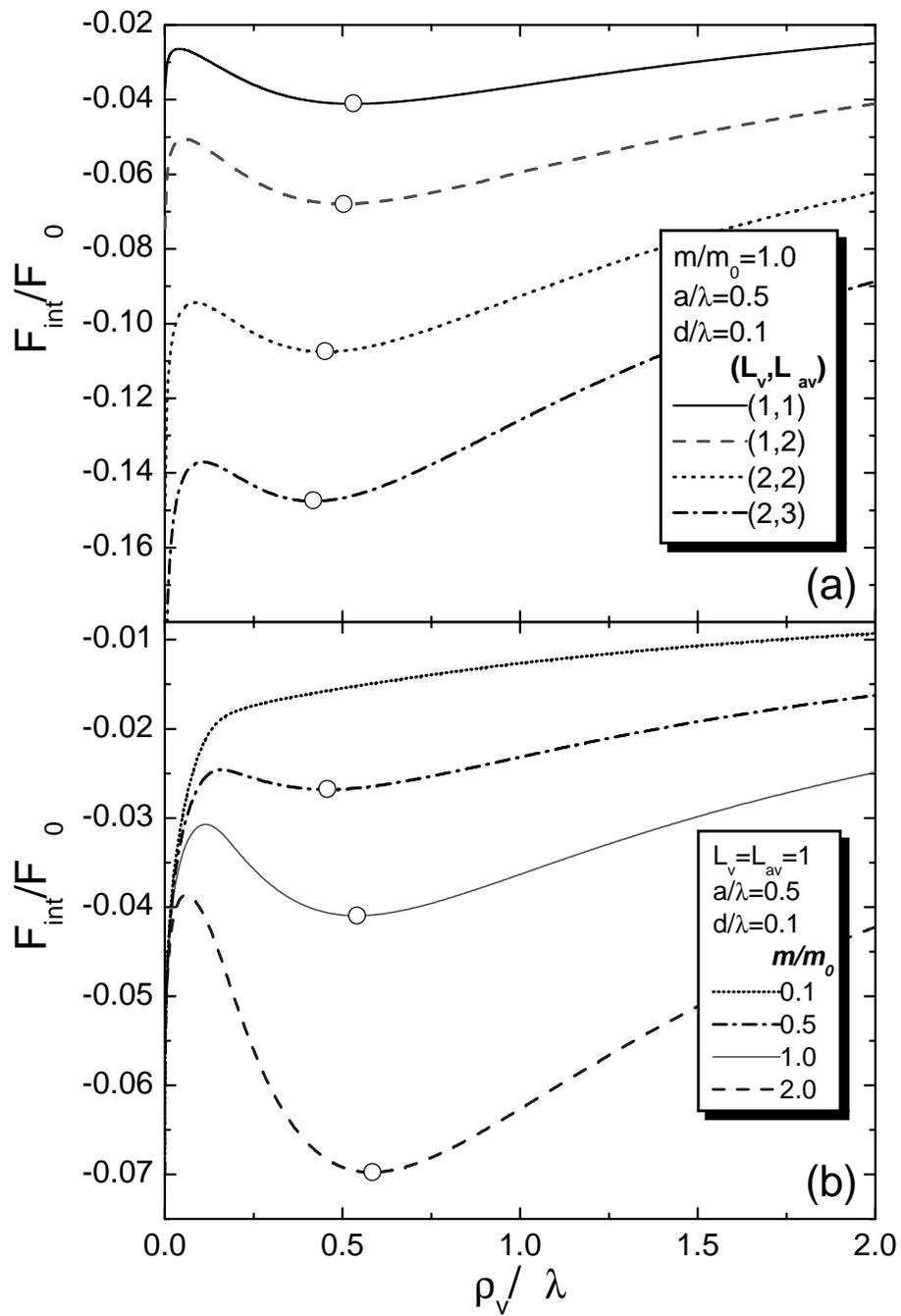
\begin{figure}[tbp]
\caption{The interaction energy of a vortex-antivortex pair and an in-plane
MD, as function of the position of the vortex ($\vec{\protect\rho}_{av}=-%
\vec{\protect\rho}_{v}$), along the direction of the magnetization of the
dipole for: a) different values of the vorticity of the vortex and the
antivortex, and b) for single-quantized vortices but different value of the
magnetization of the dipole. The open circles denote the local minimum of
the interaction energy.}
\label{fig7}
\end{figure}
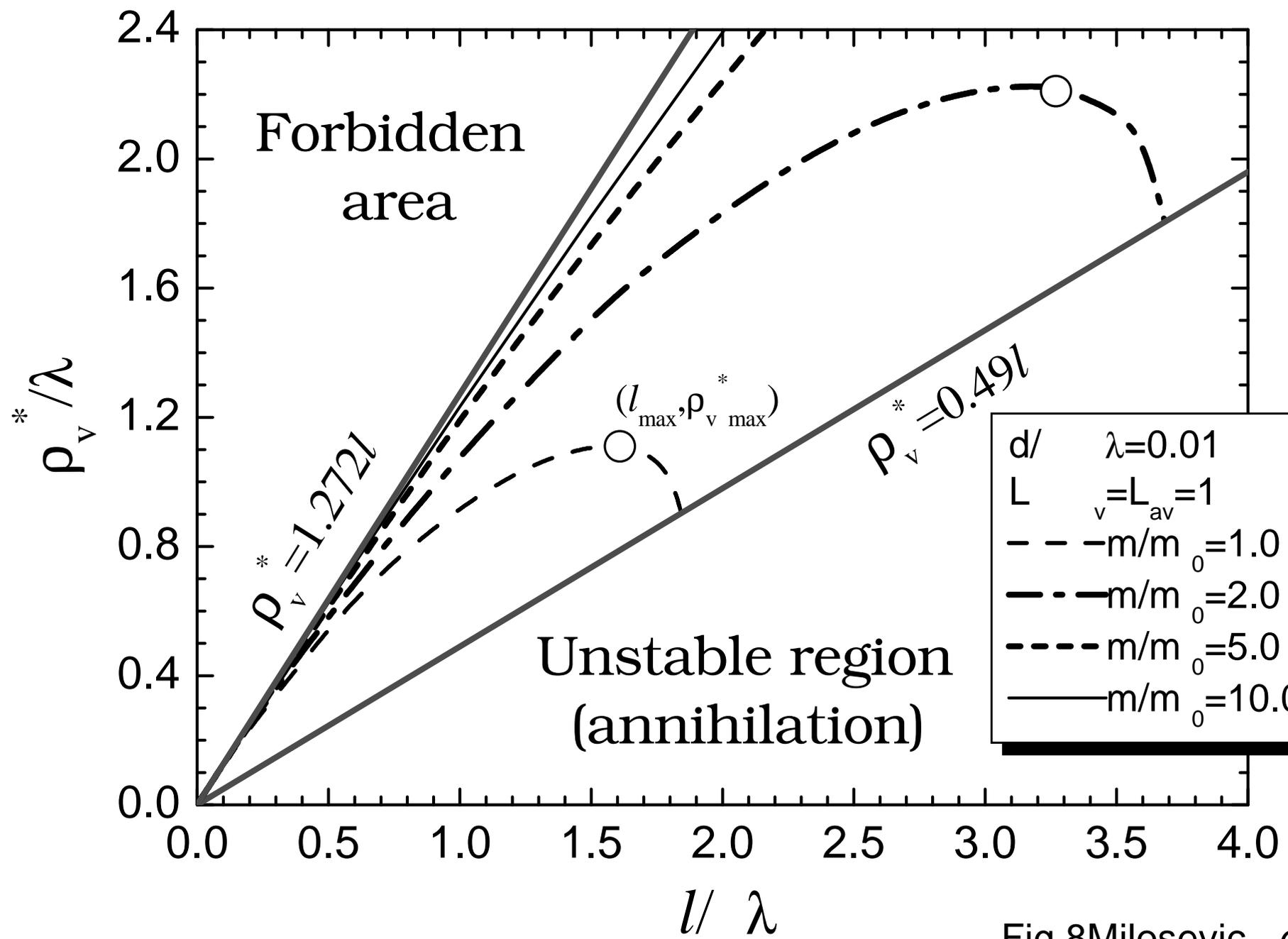
\begin{figure}[tbp]
\caption{The position of the vortex-antivortex pair ($\protect\rho %
_{v}^{\ast }$) as function of the distance between the in-plane magnetized
dipole and the SC surface ($l=a-d/2$), for different values of the
magnetization of the dipole.}
\label{fig8}
\end{figure}
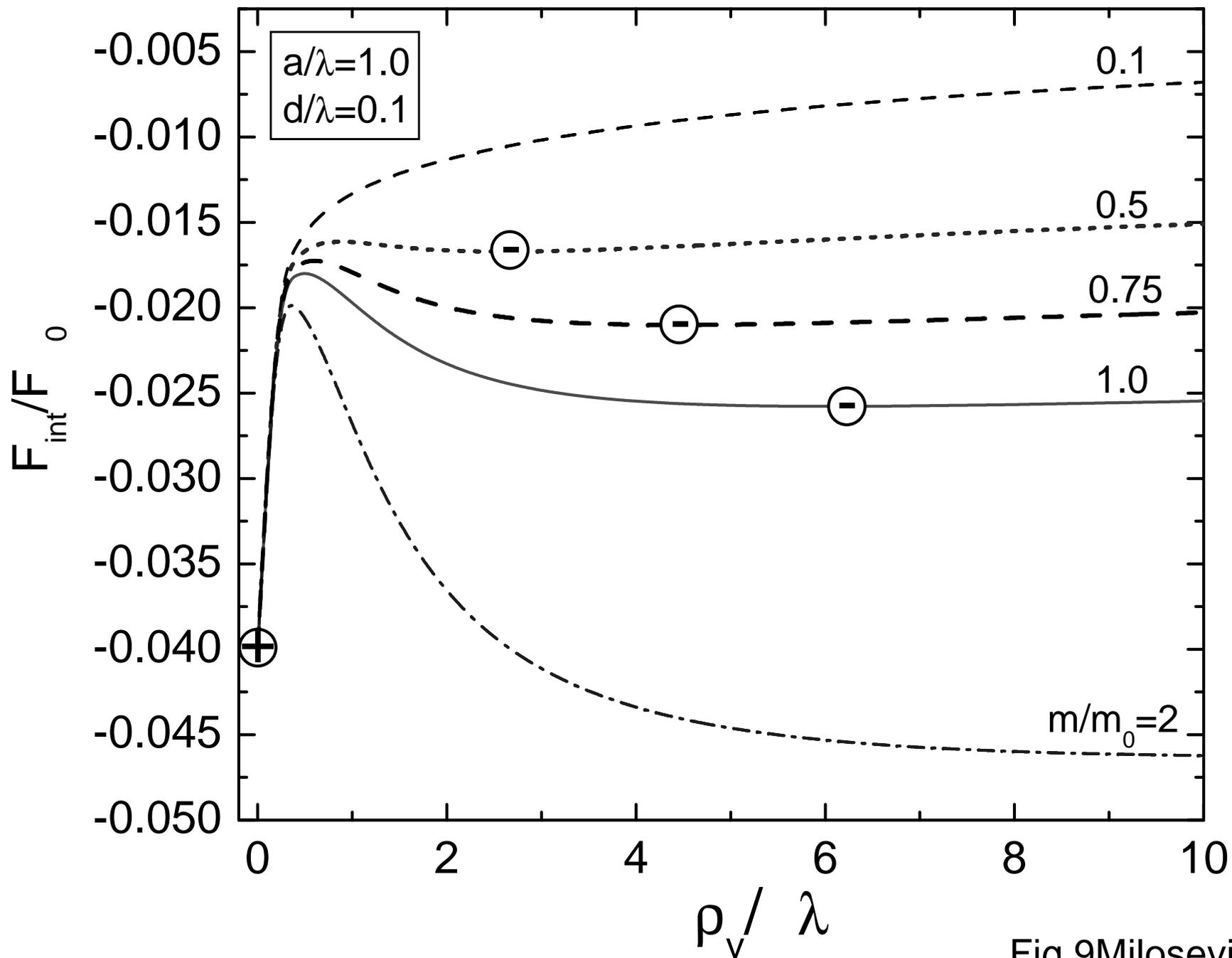
\begin{figure}[tbp]
\caption{The total interaction energy between an out-of-plane magnetic
dipole (MD), the vortex (+) under it and the antivortex (-) at a distance $%
\protect\rho _{v}$. This illustrates vortex-antivortex pair stabilization by
the short range interaction with the MD.}
\label{fig9}
\end{figure}
\begin{figure}[tbp]
\caption{The energetically favorable position of the vortex (+) and the
antivortex (-) in the case of $N=3$ pairs placed symmetrically around the
out-of-plane magnetic dipole (MD). The inset illustrates the configuration.}
\label{fig10}
\end{figure}
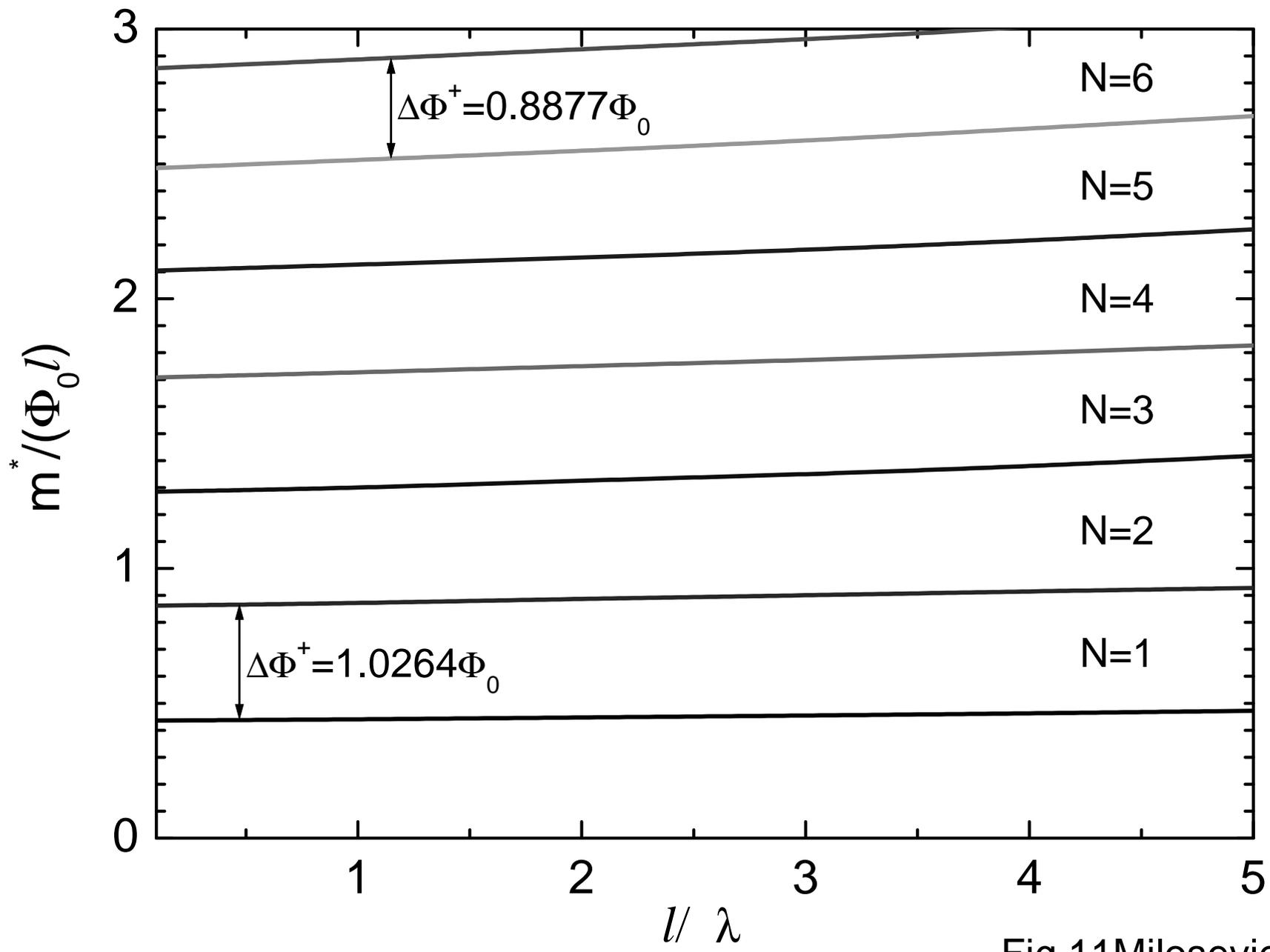
\begin{figure}[tbp]
\caption{The stability phase diagram for $N$ vortex-antivortex pairs around
the out-of-plane magnetized dipole, as function of the distance between the
dipole and the SC surface ($l$) and the magnetic moment $m$. Solid lines
denote the critical value of the moment needed for the stability of such
vortex-antivortex configurations.}
\label{fig11}
\end{figure}
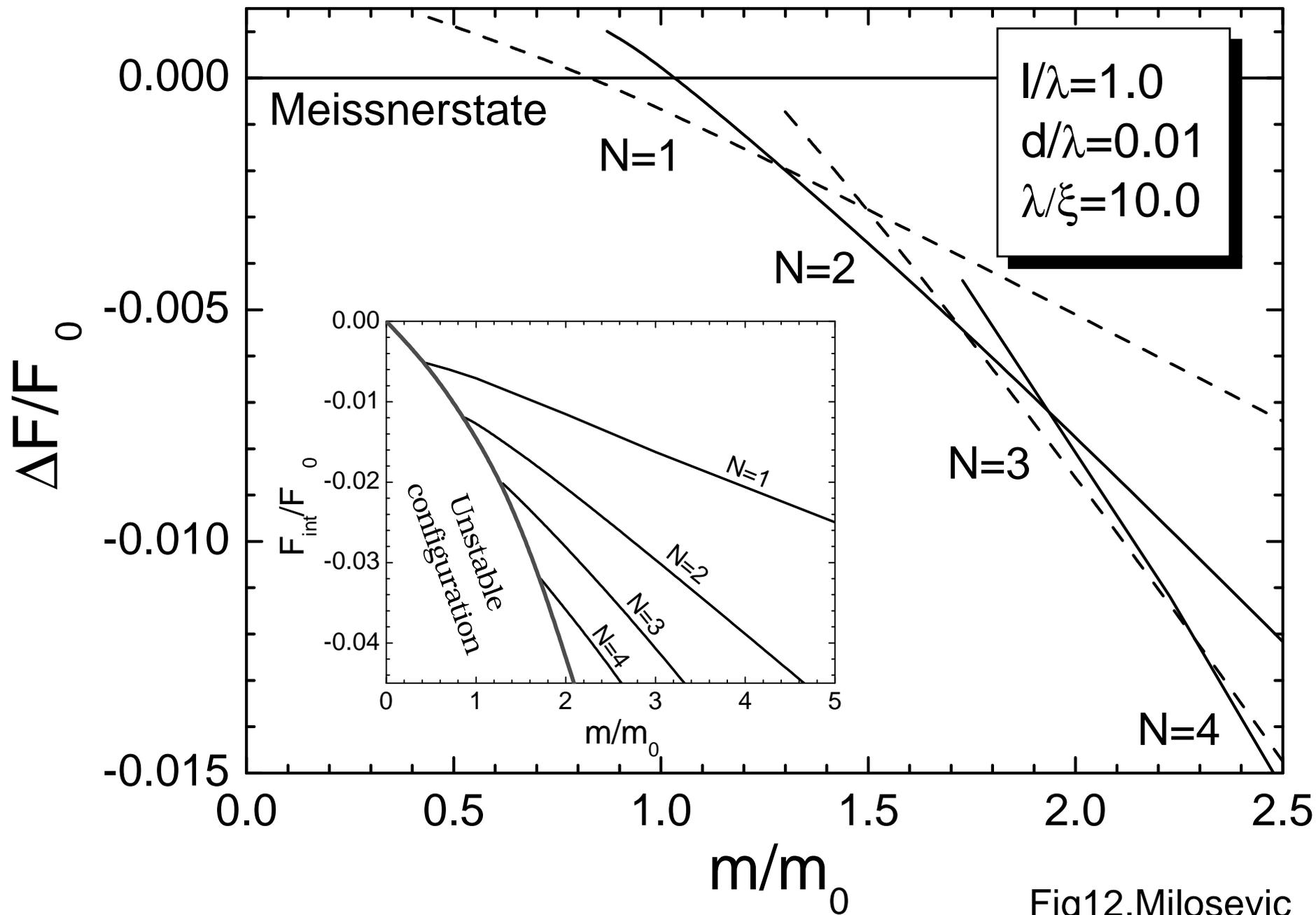
\begin{figure}[tbp]
\caption{The free energy difference between the states with $N$
vortex-antivortex pairs around the dipole and the Meissner state, for a thin
SC film with $\protect\lambda /\protect\xi =10$ and the out-of-plane dipole
positioned at $l/\protect\lambda =1.0$. Inset shows the total interaction
energy in the corresponding $N$-states.}
\label{fig12}
\end{figure}

\end{document}